\documentclass[journal]{IEEEtran}
\usepackage[ruled, vlined]{algorithm2e}
\usepackage{color,soul}
\usepackage{amsmath}

\SetKwFor{kwInit}{Initialization:}{}{}
\SetKwFor{kwFP}{Forward propagation:}{}{}
\SetKwFor{kwBP}{Backward propagation:}{}{}
\SetKw{kwTo}{to}
\SetKw{kwDownto}{downto}
\SetKw{kwEnd}{end}
\SetKw{kwDo}{do}
\SetKw{kwThen}{then}
\SetKwFor{kwForEach}{for each}{\kwDo}{\kwEnd}
\SetArgSty{textnormal}
\usepackage{comment}
\usepackage{graphicx}
\usepackage{hyperref}
\graphicspath{{./figures/}}
\usepackage{epstopdf}
\DeclareGraphicsExtensions{.eps}
\usepackage{amsmath}
\usepackage{amssymb}
\usepackage{tabularx}
\usepackage{tabu}
\usepackage{multirow}
\usepackage{hhline}
\usepackage{pbox}
\usepackage{url}
\usepackage{algpseudocode}
\usepackage{xcolor}
\usepackage{colortbl}

\newcommand{\isep}{\mathrel{{.}\,{.}}\nobreak}
\begin{document}

\title{Can We Use Split Learning on 1D CNN Models for Privacy Preserving Training?}

\author{Sharif Abuadbba, Kyuyeon Kim, Minki Kim, Chandra Thapa, Seyit A. Camtepe, Yansong Gao, Hyoungshick Kim, Surya Nepal
       
\IEEEcompsocitemizethanks{\IEEEcompsocthanksitem Sharif Abuadbba, Yansong Gao and Surya Nepal are with Data61, CSIRO  and Cyber Security CRC, Australia. e-mail: \{sharif.abuadbba, yansong.gao, surya.nepal\}@data61.csiro.au\protect\\
Chandra Thapa and Seyit A. Camtepe are with Data61, CSIRO, Australia. e-mail: \{chandra.thapa, seyit.camtepe\}@data61.csiro.au\protect\\ Professor Hyoungshick Kim, Kyuyeon Kim  and Minki Kim are with Sungkyunkwan University, SouthKorea and visting Data61, CSIRO, Australia. e-mail\{hyoung.kim, kyuyeon.kim, minki.kim\}@data61.csiro.au.\protect\\

Cite as: Sharif Abuadbba, Kyuyeon Kim, Minki Kim, Chandra Thapa, Seyit A. Camtepe, Yansong Gao, Hyoungshick Kim, Surya Nepal, ‘Can We Use Split Learning on 1D CNN Models for Privacy Preserving Training?’, The 15th ACM ASIA Conference on Computer and Communications Security (ACM ASIACCS 2020), Taipei, Taiwan, from Oct 5th to Oct 9th, 2020.}
\thanks{}
}

\markboth{}%
{Shell \MakeLowercase{\textit{et al.}}: Bare Demo of IEEEtran.cls for Computer Society Journals}

\newcommand{\sharif}[1]{\textcolor{blue}{Sharif: #1}}

\IEEEtitleabstractindextext{%
\begin{abstract}
A new collaborative learning, called \emph{split learning}, was recently introduced, aiming to protect user data privacy without revealing raw input data to a server. It collaboratively runs a deep neural network model where the model is split into two parts, one for the client and the other for the server. Therefore, the server has no direct access to raw data processed at the client. Until now, the split learning is believed to be a promising approach to protect the client's raw data; for example, the client's data was protected in healthcare image applications using 2D convolutional neural network (CNN) models. However, it is still unclear whether the split learning can be applied to other deep learning models, in particular, 1D CNN. 

In this paper, we examine whether split learning can be used to perform privacy-preserving training for 1D CNN models. To answer this, we first design and implement an 1D CNN model under split learning and validate its efficacy in detecting heart abnormalities using medical ECG data. We observed that the 1D CNN model under split learning can achieve the same accuracy of 98.9\% like the original (non-split) model. However, our evaluation demonstrates that split learning may fail to protect the raw data privacy on 1D CNN models.
To address the observed privacy leakage in split learning, we adopt two privacy leakage mitigation techniques: 1) adding more hidden layers to the client side and 2) applying differential privacy. Although those mitigation techniques are helpful in reducing privacy leakage, they have a significant impact on model accuracy. Hence, based on those results, we conclude that split learning alone would not be sufficient to maintain the confidentiality of raw sequential data in 1D CNN models.
\end{abstract}

}

\maketitle

\IEEEdisplaynontitleabstractindextext

\IEEEpeerreviewmaketitle

\IEEEraisesectionheading{}

\section{Introduction}
Deep learning has been successfully applied to many applications, including genomics~\cite{health} and healthcare systems \cite{genomics}. In such health applications, those models monitor patients' status effectively and detect their disease earlier. 
To achieve high accuracy of deep learning models, they need to be trained with sufficient data collected from a wide range of institutions \cite{gupta2018distributed}. However, sharing raw data, especially in health applications, may raise privacy concerns that violate certain rules such as reusing the data indiscriminately and risk-agnostic data processing~\cite{gdprviolations} required by General Data Protection Regulation (GDPR) \cite{gdpr} and HIPAA \cite{hipaa2003}.

In 2018, Otkrist et al.~\cite{gupta2018distributed} introduced a new collaborative learning technique, called \emph{split learning}, to protect user privacy by allowing training without sharing users' raw data to the server that runs a deep neural network (DNN) model \cite{vepakomma2018split, vepakomma2018nopeak}. Generally, split learning divides the DNN layers into two parts (A and B) between client and server. The client, who owns the raw data, trains part A that consists of the first few layers using forward propagation and only sends their activated outputs from the split layer (the last layer of the part A) to the server. After receiving the activated outputs from the client, the server performs the forward training with those outputs on part B. Next, the server runs the backward propagation on part B and only sends back the gradients of the activated outputs of the split layer (first layer of part B) to the client to complete the backward propagation on part A. This process continues until the model is converged. 

Goals  of split learning are: 1) the raw data is no longer required to be shared with the server, 2) the model classification accuracy is comparable to the non-split model~\cite{gupta2018distributed}, and 3) reducing the computational overhead of the client who only needs to run a few layers rather than the whole model. To date, the effectiveness of split learning has been validated in vision domains such as the medical image classification problem via a 2D convolutional neural network (CNN)~\cite{vepakomma2018nopeak}. However, health data includes not only images but also sequential/time-series data such as ECG signals.  


As a first study towards exploring the feasibility of split learning to deal with sequential data, we adopt an 1D CNN model for detecting heart abnormalities using ECG signals that are collected from electrodes attached to human skin \cite{garcia2019arrhythmia} as a case study. Recently, several 1D CNN models were introduced to classify sequential data, including biomedical ECG signals \cite{kiranyaz20191d,kiranyaz2015real, li2017classification}. Considering the fact that the exposure of raw ECG data would raise privacy concerns because ECG signals can reveal people's disease status and also be used to identify people uniquely~\cite{sellami2017ecg}, it is crucial  to protect the privacy of raw data for 1D CNN models. Split learning would be a promising candidate to fulfill this privacy requirement.

This work is dedicated to investigating the answers to the following two {\bf research questions (RQs):} 

\vspace{0.1cm}
\noindent\textit{\textbf{RQ 1: Can split learning be applied to deal with sequential or time-series data in particularly using 1D CNN to achieve comparable model accuracy as that of trained on centralized data?}}

To answer RQ 1, we first investigate the applicability of split learning to 1D CNN models to deal with sequential data. 
To the best of our knowledge, this is the first elaborated study on split learning using 1D CNN models, where we confirm that split learning is applicable  to sequential data. 

\vspace{0.1cm}
\noindent\textit{\textbf{RQ 2: Can split learning be used to protect privacy in sequential data trained using 1D CNN?}}

Then, we focus on understanding the privacy leakage of split learning to answer RQ 2. We find that the impact of split learning was rather limited to reduce privacy leakage.


\vspace{0.2cm}
Correspondingly, we have made the following contributions:
\begin{itemize}
    \item We implement\footnote{~\url{https://github.com/SharifAbuadbba/split-learning-1D}} split learning on 1D CNN model and apply it for time-series sequential data exemplified by using ECG signals to detect heart abnormalities, where sharing medical data with other party is inherently avoided but can still achieve the same accuracy of the non-split model. 
    \item We propose a privacy assessment framework for CNN models employing split learning, with three metrics: visual invertibility, distance correlation \cite{wang2018distance}, and Dynamic Time Warping (DTW) \cite{senin2008dtw}. This is to answer RQ 2. We observed that direct application of split learning into 1D CNN has a high privacy leakage in the applications with sensitive data such as ECG signals.
    
    \item To address the shortcoming of direct application of split learning into 1D CNN, we apply two countermeasures: i) increasing the number of convolutional layers of a CNN model split at the client and ii) exploiting differential privacy. The results suggest that although these techniques seem helpful to reduce privacy leakage, they have a significant impact on the accuracy of the model.
\end{itemize}

The rest of the paper is organized as follows: Section \ref{sec:background} provides background information about CNN, split learning, and privacy issues in using deep neural network models on cloud services. The design and implementation of split learning on 1D CNN are detailed in Section \ref{sec:designsplit1dCNN}. Section~\ref{sec:privacyanalysis} analyzes the privacy leakage question on 1D CNN models under split learning based on our identified threat model. Section \ref{sec:mitigation} discusses the possibility of two mitigation techniques and evaluates them. Section \ref{sec:discussion} discusses our findings and future work. Section \ref{sec:relatedwork} presents the related work, followed by the conclusion in Section \ref{sec:conclusion}.

\section{Background}\label{sec:background}
This section provides the necessary information to understand our work. It includes an 1D convolutional neural network, split learning technique, and privacy issues in the field of machine learning.

\subsection{1D Convolutional Neural Network (CNN)}
A CNN is a part of broader machine learning methods based on artificial neural networks where input feature extraction is performed automatically \cite{cnn:schmidhuber2015deep}.  A 1D CNN for classification problem can be depicted as a mapping function $f_\Theta: \mathbb{R}^{n\times c}\rightarrow\mathbb{R}^m$ that maps an input  $x \in \mathbb{R}^{n\times c}$ to an output  $\hat{y}\in\mathbb{R}^m$ based on the calculated parameters $\Theta$, where $n$ is the length of input vector, $c$ is the number of input channels, and $m$ is the number of classes. For example, let us assume $x$ is an ECG sample taken from a patient that has to be classified by $f_\Theta$  into $\hat{y}$ as a vector of probabilities corresponding to  5 different types of heartbeat diseases.  The output with the highest value, $\textup{arg max}_{i\in{\{1\isep 5\}}}\hat{y_i}$, is a final prediction from the model in which $x$ is most likely to be, e.g., `$A$' (atrial premature contraction).

A CNN  is constructed with $L$ hidden layers. Each layer $l$, $ l\in \{1\isep L\}$, has $n_l$ neurons, and activated output $a^{(l)}$. The vector consists of values of each neuron of that layer, and it is computed in a feed-forward propagation manner as follows:

\begin{equation}\label{eq:activations}
a^{(l)} = g^{(l)}(w^{(l)}a^{(l-1)}+b^{(l)})\;\;\;  \forall\ l \in \{1\isep L\}
\end{equation} 
where  $g^{(l)}: \mathbb{R}^n \rightarrow \mathbb{R}^n$ is a non-linear activation function of layer $l$ that ensures only crucial neurons to be fired (i.e., $>0$) and forwards its output as an input to the next layer.  $w^{(l)}\in \mathbb{R}^{n_l \times n_{l-1}}$ is the \textit{weights} and  $b^{(l)} \in \mathbb{R}^{n_l}$ is the \textit{biases}; both of them are learned during training. The CNN output of the $L$ layer, i.e., the last hidden layer, is a function which  can be calculated as  $\hat{y_i}=a^{(L)}=g^{(L)}(w^{(L+1)}a^{(L)} + b^{(L+1)})$. After performing forward propagation reaching the output $y_i$, the difference between the ground truth label $y_i$  and predicted  $\hat{y_i}$ is calculated as the loss $E_i =(y_i - \hat{y_i})^2$, in case of using squared loss. Then, the contribution of every $w$ towards this loss is calculated in a backpropagation way.
For that, the partial derivatives of the loss with respect to each individual weight is calculated. As an example, we present the calculation with respect to a single weight  $w^{(L)}_{12}$ that connects node $2$ in layer $L-1$ to node $1$ in Layer  $L$ as follows:

\begin{equation} \label{eq:partial0}
\frac{\partial E_i}{\partial w^{(L)}_{12}} =(\frac{\partial E_i}{\partial a^{(L)}_{1}})(\frac{\partial a^{(L)}_{1}}{\partial z^{(L)}_{1}})(\frac{\partial z^{(L)}_{1}}{\partial w^{(L)}_{12}})
\end{equation} 

\begin{equation}\label{eq:partial}
=2(a^{(L)}_{1}-y_i)(g^{(L)}{'}(z^{(L)}_1))(a_2^{(L-1)})
\end{equation}
where  $g^{(L)}$ is the activation function of layer $L$ and  $z^{(L)}$ is the input for that neuron.

\begin{figure}[!h]
	\centerline
	{\includegraphics[scale=0.53]{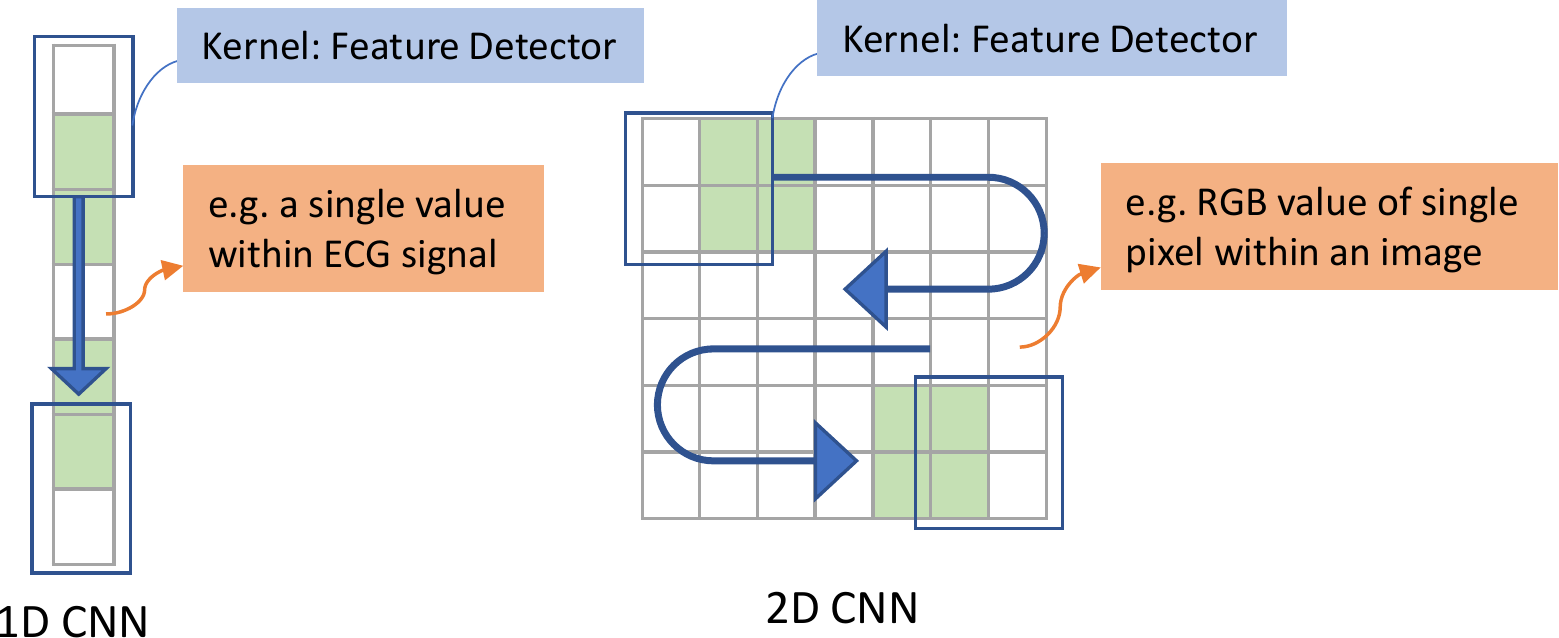}}
	\caption{1D CNN vs. 2D CNN in feature detection. The shape of convolution filter is a \textbf{\textit{vector}} form in 1D CNN while a \textbf{\textit{2D matrix}} form in 2D CNN.}
	\label{fig:1dvs2d}
\end{figure}

The widely used CNN models are 1D and 2D CNN. The 2D CNN is  popular in image recognition to extract features from 2D images. The 1D CNN also recently produced several promising results in extracting features from sequential time-series data \cite{li2017classification}. Both types are using similar steps as in Equation \eqref{eq:activations}, \eqref{eq:partial0} and \eqref{eq:partial}, but the major difference is the structure of the input data and how the convolution filter, also known as a kernel or feature detector, moves across the data for feature extraction. The shape of convolution filter (kernel) is a  {\it vector} form in 1D CNN and usually {\it 2D matrix} form in 2D CNN, as shown in Fig. \ref{fig:1dvs2d}.

\subsection{Split Learning}
Split learning is a distributed deep learning technique that splits a CNN into two parts; the first part is provided to the client and the second part to the server. Both client and server collaboratively train the split model without accessing each other's part. To perform the split learning for 2D CNN models, several networks such as LeNet, VGG, and AlexNet were considered and validated~\cite{gupta2018distributed}. 
\subsection{Privacy Preserving Machine Learning}\label{sec:threatmodel}

Cloud servers or major model providers have been popularly used for collecting and processing data. However, users often have privacy concerns when their sensitive data is processed and stored at cloud servers~\cite{cloudserversurvey}. In practice, user data on the cloud server can be misused to identify individuals even though their explicit identifier information is not provided. For example, a previous study~\cite{sellami2017ecg} showed ECG signals can be used to uniquely identify individuals. Perhaps, identification of individuals may violate important privacy rules such as reusing the data indiscriminately, and risk-agnostic data processing~\cite{gdprviolations} required by regulations such as GDPR in Europe~\cite{gdpr}.

In this regard, the privacy-preserving machine learning technique through distributed learning, such as federated learning~\cite{konevcny2016federated} and split learning~\cite{vepakomma2018split}, is promising. Split learning as the focus of this work aims to reduce the privacy leakage of sensitive localized data by splitting the network during training---allowing raw data being remained in the data owner (i.e., client).  However, there is a possibility of privacy leakage from the information sent from the client during the machine learning process. Precisely, the privacy leakage is that given the activation at the split layer $l$, i.e., $a^{(l)}$, how much one (the server) infers about the training data $x$. 




\section{Design and Implementation of the Split 1D CNN}\label{sec:designsplit1dCNN}
In this section, we design and implement the split 1D CNN to answer the RQ 1: 
\noindent\textit{\textbf{Can split learning be applied to deal with sequential/time-series data in particularly using 1D CNN to achieve comparable model accuracy as that of trained on centralized data?}}

We first introduce the 1D CNN ECG classification models \cite{kiranyaz2015real, li2017classification} that we reproduced. We then detail our implementation of splitting the 1D CNN model. Consequently, we validate that the split 1D CNN is able to achieve the same model accuracy of the non-split 1D CNN, where our RQ 1 is  answered.

\subsection{1D CNN for ECG Signal Classification}

In this section, we describe the non-split version of 1D CNN ECG classification models, which were recently introduced to classify ECG signals~\cite{li2017classification, kiranyaz2015real, kiranyaz2017personalized, wu2018comparison,yildirim2018arrhythmia}. 
We chose two 1D CNN model architectures given in \cite{kiranyaz2015real} and \cite{li2017classification} as they are most recent and showed the best-achieved accuracy. Both works \cite{kiranyaz2015real,li2017classification} aim to classify ECG signals into 5 classes with less than or equal to 5 layers of 1D CNN. For the model architecture in \cite{kiranyaz2015real}, it has three 1D CNN layers and two fully connected layers, exhibiting about 96.6\% accuracy. In \cite{li2017classification}, only two 1D convolutional layers are used with two fully connected layers, demonstrating about 97.5\% accuracy. We first implement these original non-split model from those two studies and then implement them using split learning to validate consistency in the model accuracy.

\subsubsection{\textbf{ECG Dataset and Preprocessing.}}

We use MIT-BIH arrhythmia \cite{moody2001impact} which is a popular dataset for ECG signal classification or arrhythmia diagnosis detection models. Arrhythmia is short for Abnormal Heart Rythm, which is an indication of various heart diseases. Following the models \cite{kiranyaz2015real, li2017classification}, we collect 26,490 samples in total which represents 5 types of heartbeat as classification targets: $N$ (normal beat), $L$ (left bundle branch block), $R$ (right bundle branch block), $A$ (atrial premature contraction), $V$ (ventricular premature contraction). We normalize these samples and remove the noise before feeding them to the 1D CNN as shown in Fig. \ref{fig:preprocess}. We explain the detailed preprocessing steps in  Appendix \ref{sec:appendixA}.
\begin{figure}[!h]
	\centerline
	{\includegraphics[scale=0.34]{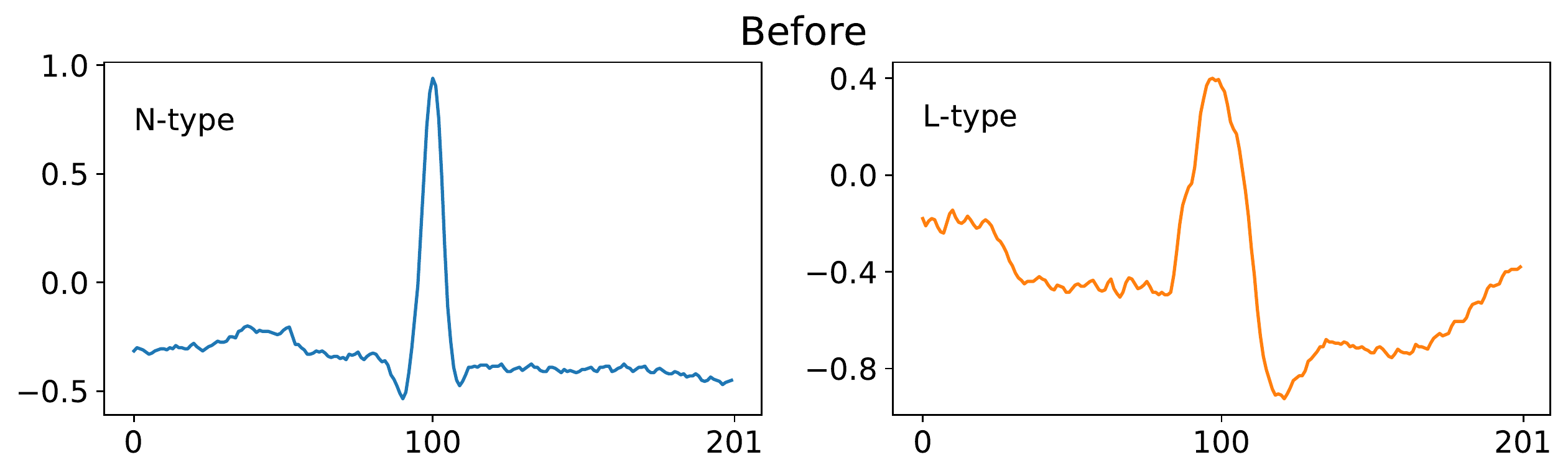}}
	\centerline
	{\includegraphics[scale=0.34]{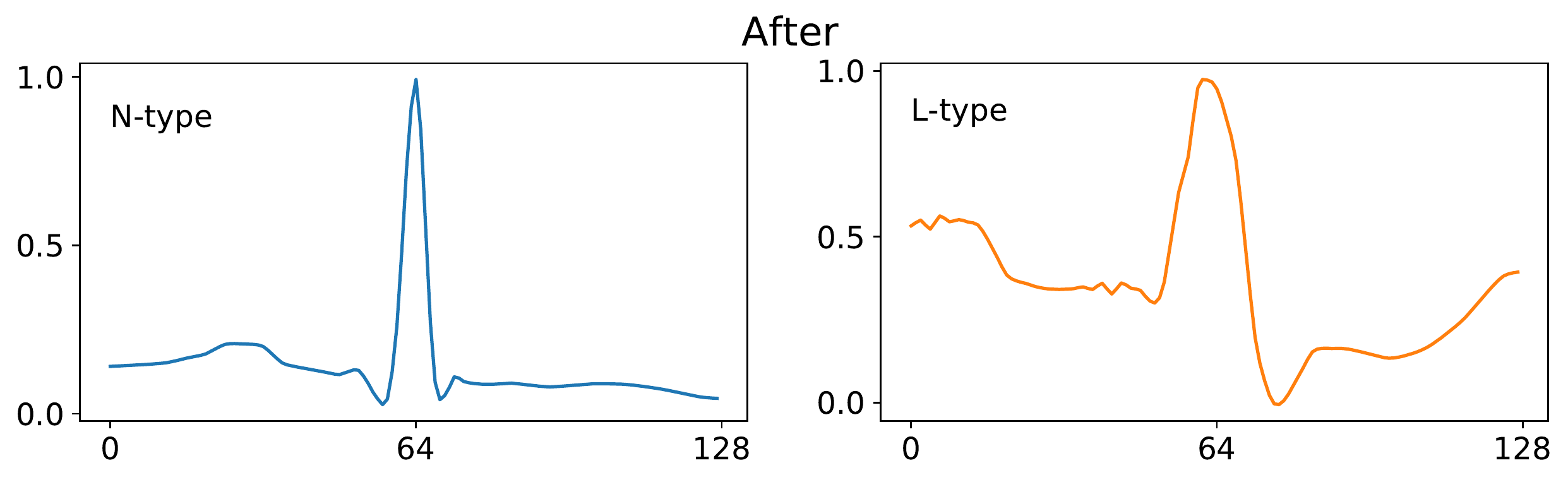}}
	\caption{ECG signals before and after preprocessing.}
	\label{fig:preprocess}
\end{figure} 

\begin{figure}[!h]
	\centerline
	{\includegraphics[scale=0.38]{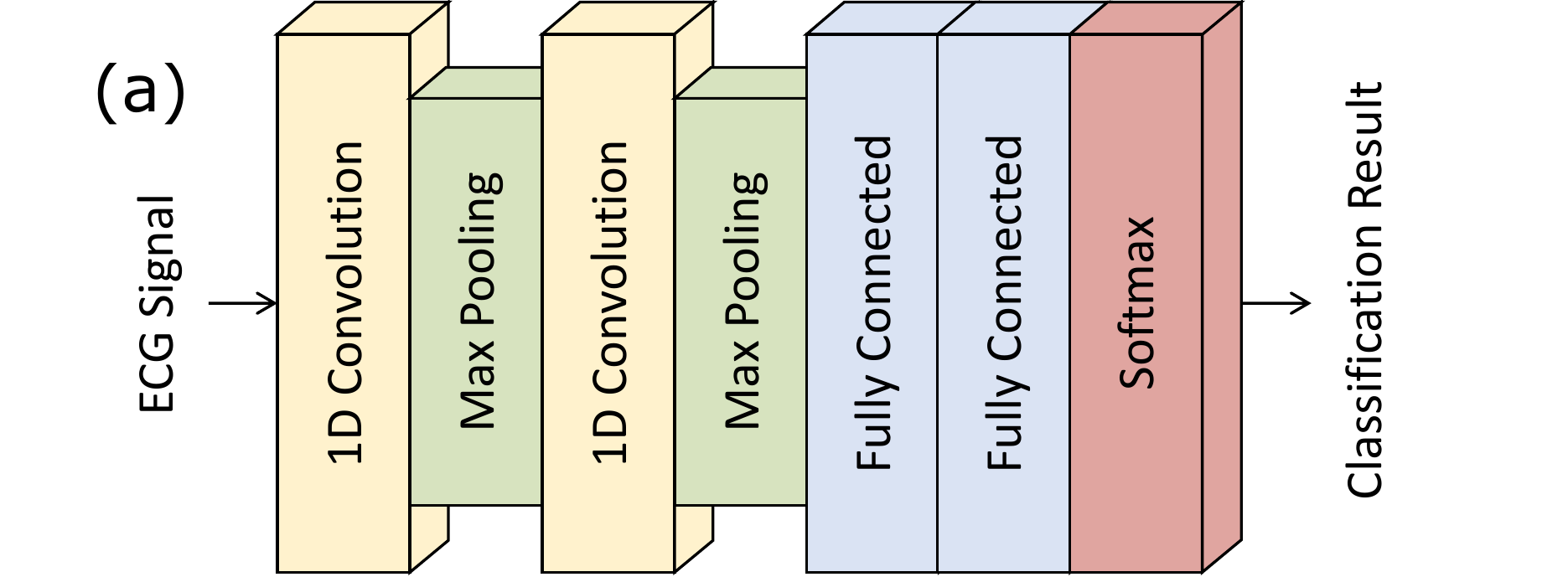}}
	\centerline
	\centerline
	{\includegraphics[scale=0.38]{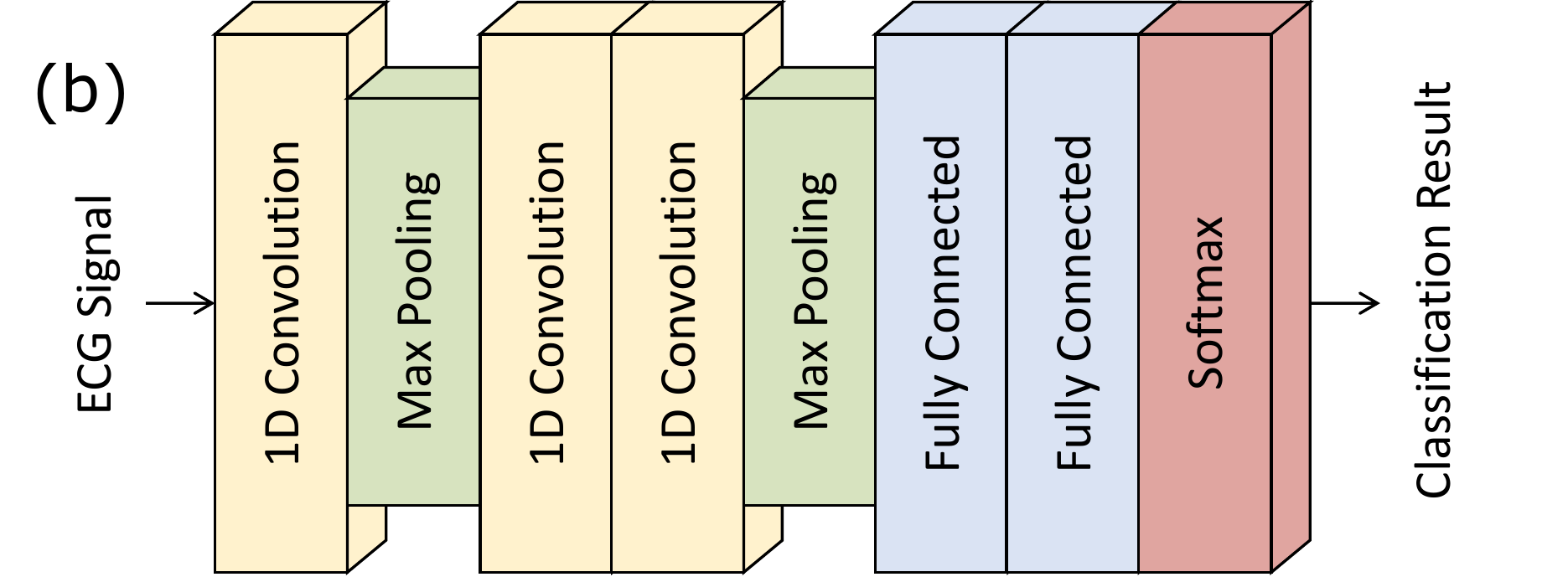}}
	\caption{1D CNN for ECG signal classification using (a) two or (b) three 1D convolutional layers.}
	\label{fig:ecgmodel}
\end{figure}

\subsubsection{\textbf{1D CNN Model Architecture.}}\label{sec3.1.2}
The first 1D CNN \cite{li2017classification} model architecture we adopted is illustrated in Fig.~\ref{fig:ecgmodel} (a). Specifically, each convolutional layer has 16 filters: the size of the filter used for the first convolutional layer is 7, and 5 for the rest. Zero padding is applied before each convolution operation.  
Rather than ReLU, Leaky ReLU is chosen as an activation function of hidden layers to prevent the dying ReLU problem. Softmax is used for the activation function of the last fully connected layer. We call this model `two-layer model'. The second 1D CNN model adopted by us~\cite{kiranyaz2015real} is with three 1D convolutional layers (see Fig.~\ref{fig:ecgmodel} (b)), termed as `three-layer model'. Parameter settings are similar to the two-layer model.


\begin{table}[!h] 
\caption{ECG dataset specifications.}
\label{tb:dataset}
\renewcommand{\arraystretch}{1.5}
\begin{tabular}{c|c|c|c|c|c|c}
\hline
Class                   & N     & L     & R     & A     & V     & Total  \\ \hline \hline
Train                  & 3,000 & 3,000 & 3,000 & 1,245 & 3,000 & 13,245 \\ \hline
Test                   & 3,000 & 3,000 & 3,000 & 1,245 & 3,000 & 13,245 \\ \hline
Total                  & 6,000 & 6,000 & 6,000 & 2,490 & 6,000 & 26,490 \\ \hline \hline
\end{tabular}
\renewcommand{\arraystretch}{1}
\end{table}
\subsubsection{\textbf{Training Result.}}
Table \ref{tb:dataset} shows our training and testing dataset distribution which follows a similar proportion setting for each of 5 classes as in the previous work~\cite{li2017classification}. 
Both two-layer model and three-layer model are trained with 400 epochs, respectively. The learning rate is set to 0.001. Adam optimizer with a batch size of 32 is set.

\begin{figure*}[!ht]
	\centerline
	{\includegraphics[scale=0.29]{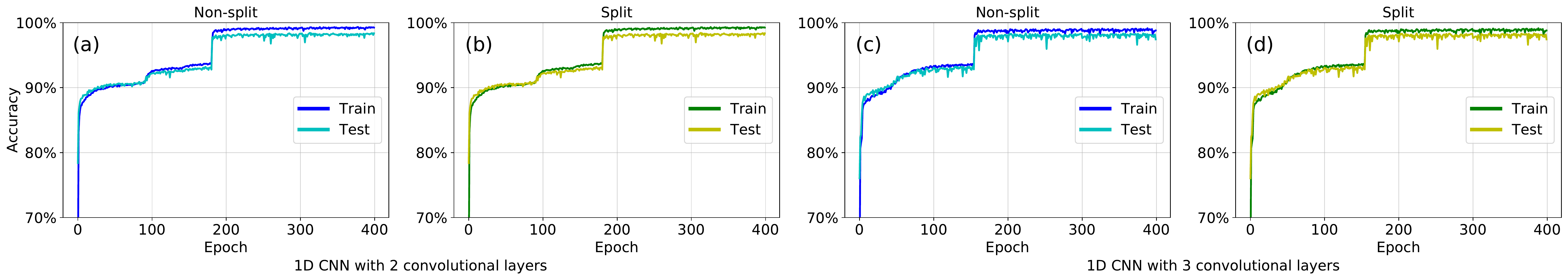}}
	\caption{Accuracy over the training of split and non-split 1D CNN models. Notably, we have set the exact same initial weights through using the equivalent random seed, for both split and non-split tests in this example.} 
	\label{fig:loss_acc_split_vs_nonsplit}
\end{figure*}

We measure the accuracy of the models on the test set, after each epoch. As shown in Fig.  \ref{fig:loss_acc_split_vs_nonsplit} (non-split), both models' accuracy converged to 98\% around or less than 200 epochs. The accuracy is  not necessarily improved after reaching the optimal value. The test accuracy of two-layer model is 98.9\%, which is also similar to the test accuracy of three-layer model. Note that the original work by Dan et al. \cite{li2017classification} that we follow shows a 97.5\% accuracy in a similar setting. In other words, we have successfully reproduced their work and slightly improved the accuracy, where the improvement attributes to the small modifications in hyper-parameters and data preprocessing steps, as detailed in Appendix \ref{sec:appendixA}.


\subsection{Splitting 1D CNN}
We now split the above 1D CNN models. We focus on two parties in the split learning setting: the client and the server. As we show later, the leakage is resulted from passing the forward activation to the server, which is irrelevant to the number of participated clients.

\begin{figure}[!h]
	\centerline
	{\includegraphics[scale=0.7]{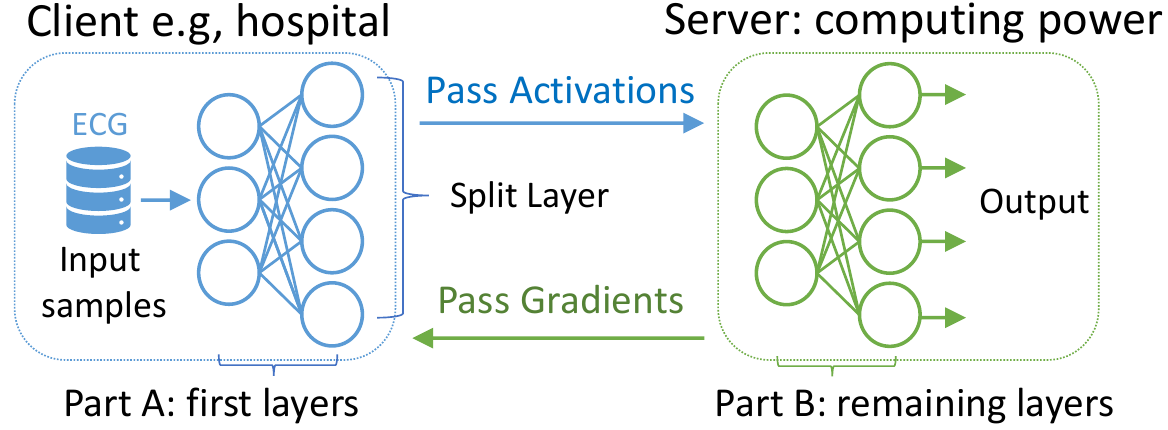}}
	\caption{Split learning overview.}
	\label{fig:architecture}
\end{figure} 

The 1D CNN model is split into two parts, A and B as shown in Fig. \ref{fig:architecture}. Activation and gradients are passed between client and server to collaboratively train the joint model.


We follow the vertical split method~\cite{gupta2018distributed,vepakomma2018split} to split 1D CNN. Those previous studies provide the conceptual process of split learning for 2D CNN models and show that model accuracy of split training is the same as that in the non-split training~\cite{gupta2018distributed,vepakomma2018split}. Other than focusing on 2D CNN models as~\cite{gupta2018distributed,vepakomma2018split}, we, herein, elaborate on 1D CNN split implementation strategies for client and server sides, respectively (see Algorithm \ref{algorithm1} and \ref{algorithm2}). Those detailed algorithms can be useful as a guideline for implementing split learning models. We also clearly specify what information is exchanged between client and server with the socket instructions in the pseudocodes in Algorithm \ref{algorithm1} and \ref{algorithm2}, which will be especially helpful for practitioners along with our source code. 

\begin{algorithm}[t]
\caption{Split learning on the client side}\label{algorithm1}
\kwInit{}{
    $s\gets$ socket initialized with port and address\\
    $s.connect(Bob)$\\
    $\phi,\eta,o,n,N\gets s.synchronize()$\\
    $\{w^{(i)}\}_{\forall i\in\{0\isep l\}}\gets$ initialize using $\phi$\\
    $\{z^{(i)}\}_{\forall i\in\{0\isep l\}}, \{a^{(i)}\}_{\forall i\in\{0\isep l\} }\gets\emptyset$\\
    $\{\frac{\partial E}{\partial z^{(i)}}\}_{\forall i\in\{0\isep l\}}, \{\frac{\partial E}{\partial a^{(i)}}\}_{\forall i\in\{0\isep l\}}\gets\emptyset$
}
\kwForEach{\text{batch} $(x, y)$ \text{generated from} $\mathbb{D}$}{
    \kwFP{}{
        $a^{(0)}\gets x$\\
        \For{$i\gets 1$ \kwTo $l$}{
            $z^{(i)}\gets f^{(i)}(a^{(i-1)})$\\
            $a^{(i)}\gets g^{(i)}(z^{(i)})$
        }
        $s.send((a^{(l)}, y))$\\
    }
    \kwBP{}{
        $\frac{\partial E}{\partial a^{(l)}}\gets s.receive()$\\
        \For{$i\gets l$ \kwDownto $1$}{
            $\frac{\partial E}{\partial z^{(i)}}\gets \frac{\partial E}{\partial a^{(i)}}\times g^{(i)}{'}(z^{(i)})$\\
            Compute $\frac{\partial E}{\partial w^{(i)}}$ using $\frac{\partial E}{\partial z^{(i)}}$ and $a^{(i-1)}$\\
            Update $w^{(i)}$ using $\eta$, $\frac{\partial E}{\partial w^{(i)}}$, and $o$\\
            \If{$i\neq 1$}{
                $\frac{\partial E}{\partial a^{(i-1)}}\gets f^{(i)}_{\textup{T}}(\frac{\partial E}{\partial z^{(i)}})$
            }
        }
    }
}
$s.close()$
\end{algorithm}

\begin{algorithm}[t]
\caption{Split learning on the server side}\label{algorithm2}
\kwInit{}{
    $s\gets$ server socket initialized with port and address\\
    $s_A\gets s.accept(Alice)$\\
    $\phi,\eta,o,n,N\gets s_A.synchronize()$\\
    $\{w^{(i)}\}_{\forall i\in\{l+1\isep L\}}\gets$ initialize using $\phi$\\
    $\{z^{(i)}\}_{\forall i\in\{l+1\isep L\}}, \{a^{(i)}\}_{\forall i\in\{l\isep L\}}\gets\emptyset$\\
    $\{\frac{\partial E}{\partial z^{(i)}}\}_{\forall i\in\{l+1\isep L\}}, \{\frac{\partial E}{\partial a^{(i)}}\}_{\forall i\in\{l\isep L\}}\gets\emptyset$
}
\For{$i\gets 1$ \kwTo $N$}{
    \kwFP{}{
        ($a^{(l)},y)\gets s_A.receive()$\\
        \For{$i\gets l+1$ \kwTo $L$}{
            $z^{(i)}\gets f^{(i)}(a^{(i-1)})$\\
            $a^{(i)}\gets g^{(i)}(z^{(i)})$
        }
    }
    $E\gets \mathcal{L}(a^{(L)}, y)$\\
    \kwBP{}{
        Compute $\frac{\partial E}{\partial a^{(L)}}$\\
        \For{$i\gets L$ \kwDownto $l+1$}{
            $\frac{\partial E}{\partial z^{(i)}}\gets \frac{\partial E}{\partial a^{(i)}}\times g^{(i)}{'}(z^{(i)})$\\
            Compute $\frac{\partial E}{\partial w^{(i)}}$, using $\frac{\partial E}{\partial z^{(i)}}$ and $a^{(i-1)}$\\
            Update $w^{(i)}$, using $\eta$, $\frac{\partial E}{\partial w^{(i)}}$, and $o$\\
            $\frac{\partial E}{\partial a^{(i-1)}}\gets f^{(i)}_{\textup{T}}(\frac{\partial E}{\partial z^{(i)}})$
        }
        $s_A.send(\frac{\partial E}{\partial a^{(i)}})$
    }
}
$s_A.close()$
\end{algorithm}

\subsubsection{\textbf{Client.}}
Assume a model that has $L$ layers---input layer is excluded---in total. The $L$-th layer is the output layer, and the remaining layers are hidden layers. Suppose that the model is split between layer $l$ and layer $l + 1$. The client holds first $l$ layers from the layer $1$ to $l$---part A, whereas the server holds remaining layers from the layer $l + 1$ to $L$---part B. Weights in the layer $i$ are denoted as $w^{(i)}$. In addition, let $f^{(i)}$ denotes the forward propagation over the $i$-th layer, and $z^{(i)}$ denotes the output tensor just after the forward propagation in $i$-th layer. $a^{(i)}$ is denoted as the output after the activation function in layer $i$, which can be given by $a^{(i)}=g^{(i)}(z^{(i)})$, where $g^{(i)}$ is the activation function of $i$-th layer. In backpropagation, $f^{(i)}_{\textup{T}}$ denotes the function which returns the gradient of activation of layer $i-1$, using weights and gradient of $i$-th layer. Finally, when the client has the ECG raw dataset $\mathbb{D}$ to train with, the split learning process on client follows Algorithm \ref{algorithm1}.

The client first connects to the server via socket and synchronizes some train configurations. With a single batch given from $\mathbb{D}$, the client forward propagates it until the $l$-th layer and sends the activation $a^{(l)}$ from the $l$-th layer to the server. When the client receives the gradient of $\frac{\partial E}{\partial a^{(l)}}$ from the server, the backpropagation continues to the first hidden layer.

\subsubsection{\textbf{Server.}}
The server continues forward propagation after receiving activation from $l$-th layer. The server then calculates the loss between the activated output from the last layer and the label passed from the client. Let $E$ denotes the loss calculated from the loss function $\mathcal{L}$. With $E$, the server starts backpropagation until layer $l+1$. The server finally sends the gradient of the output to the client, which is $\frac{\partial E}{\partial a^{(l)}}$, to make the client continues the backpropagation. The rest of the denotations except $E$ and $\mathcal{L}$, are the same as used in Algorithm \ref{algorithm1}. 

The training flow between the client and the server is illustrated in Fig.~\ref{fig:flow} and further detailed in Appendix \ref{sec:appendixB}.


\begin{figure}[!ht]
	\centerline
	{\includegraphics[scale=0.6]{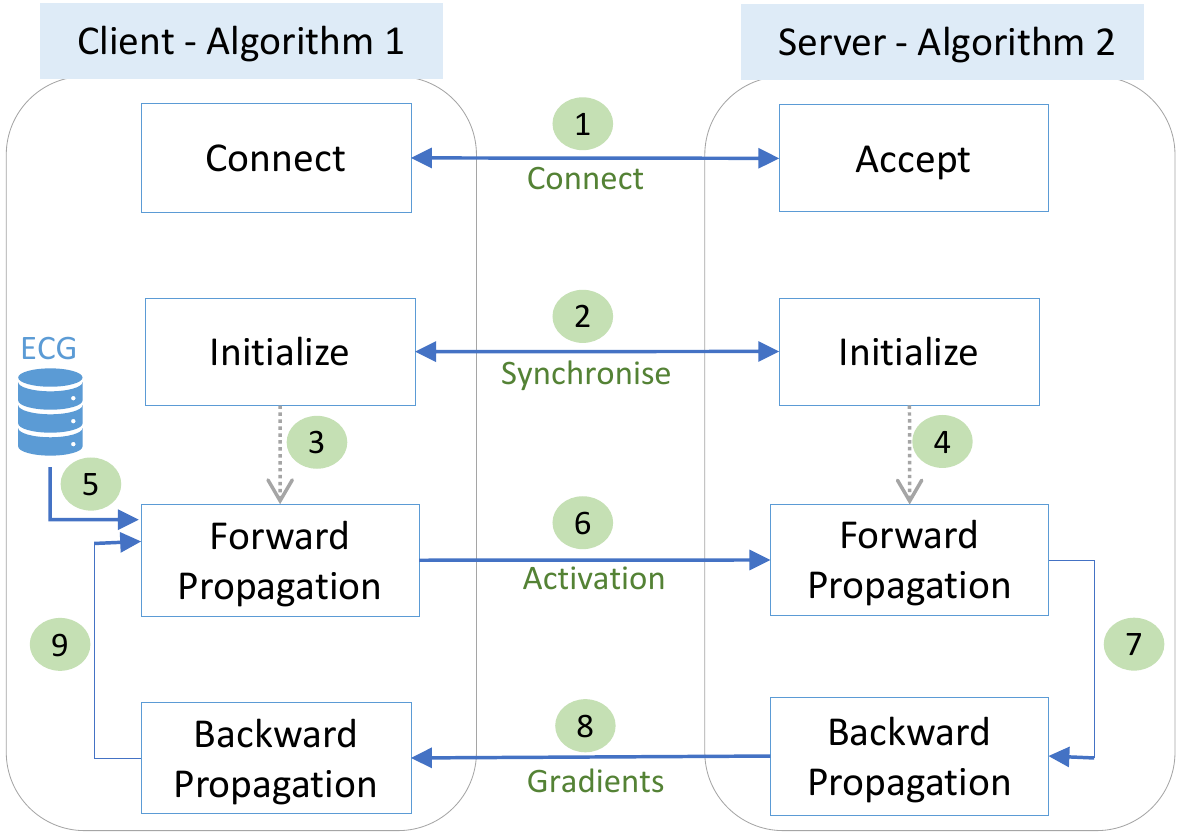}}
	\caption{Data flow between the client and the server in Algorithm \ref{algorithm1} and \ref{algorithm2}.}
	\label{fig:flow}
\end{figure} 


\subsubsection{\textbf{Influence on Performance}}
Based on the implementation above, we test the split learning on both 1D CNN model architectures: two-layer model and three-layer model. We split the former after two convolutional layers and latter after three convolutional layers. To measure and compare the influence of split learning on performance accurately, we initialize both models before and after the split with the same set of weights. Fig. \ref{fig:loss_acc_split_vs_nonsplit} depicts the exact accuracy conversion of both models before/after applying split. It is clear that the accuracy is the same. In other words, our split learning implementation has no noticeable impact on the performance of the models.

\vspace{0.2cm}
\noindent{\bf Summary:} Split learning  can be applied into 1D CNN without  the model classification accuracy degradation as demonstrated in Fig. \ref{fig:loss_acc_split_vs_nonsplit}. Therefore, RQ 1 can be is answered affirmatively. 
\section{Privacy Leakage Analysis} \label{sec:privacyanalysis}
In this section, we propose a privacy assessment framework under our identified threat model to answer the RQ 2:
\noindent\textit{\textbf{Can split learning be used to protect privacy in sequential/time-series data trained using 1D CNN?}}

This framework has three metrics: visual invertibility, distance correlation and Dynamic Time Warping. Based on these metrics, we present our empirical results validated from systematic experiments to demonstrate that it is possible to reconstruct raw data from the activation of the intermediate split layer, which indicates that  our RQ 2 is answered unfavourably.

We firstly elaborate on the considered threat model.

\subsection{Threat Model}
We consider the server is an honest-but-curious single entity adversary. It performs all its operations as specified, but curious about the raw data localized at the client. We assume that the server has no access to the client's device, and it does not target attacks on those devices. Furthermore, the server does not collude with any client. The server's goal is to reconstruct the raw data (e.g., the client's medical data) from the activated vector of the split layer, which is delivered from the client during the forward propagation. We assume that all participating clients are trusted, and they participate in the learning process provided that the raw data always remain within their custody. In the health domain, examples of trusted clients are patients, hospitals, and (health) research organizations.

\subsection{Visual Invertibility}
\begin{figure}[h]
	\centerline
	{\includegraphics[scale=0.22]{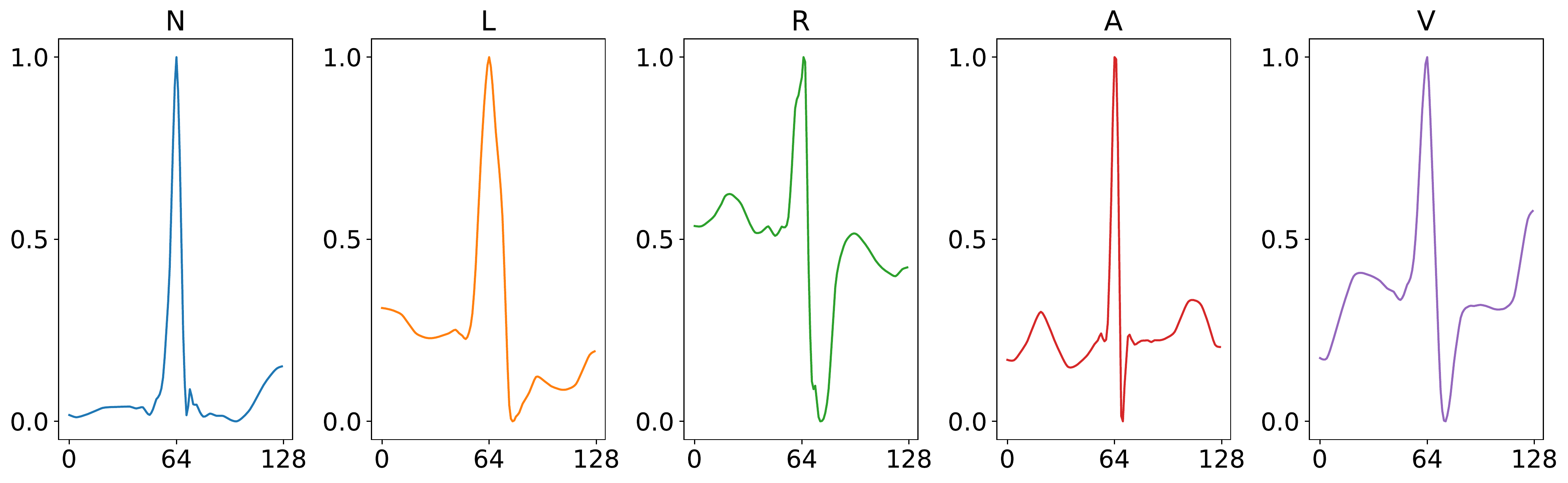}}
	\centerline
	{\includegraphics[scale=0.22]{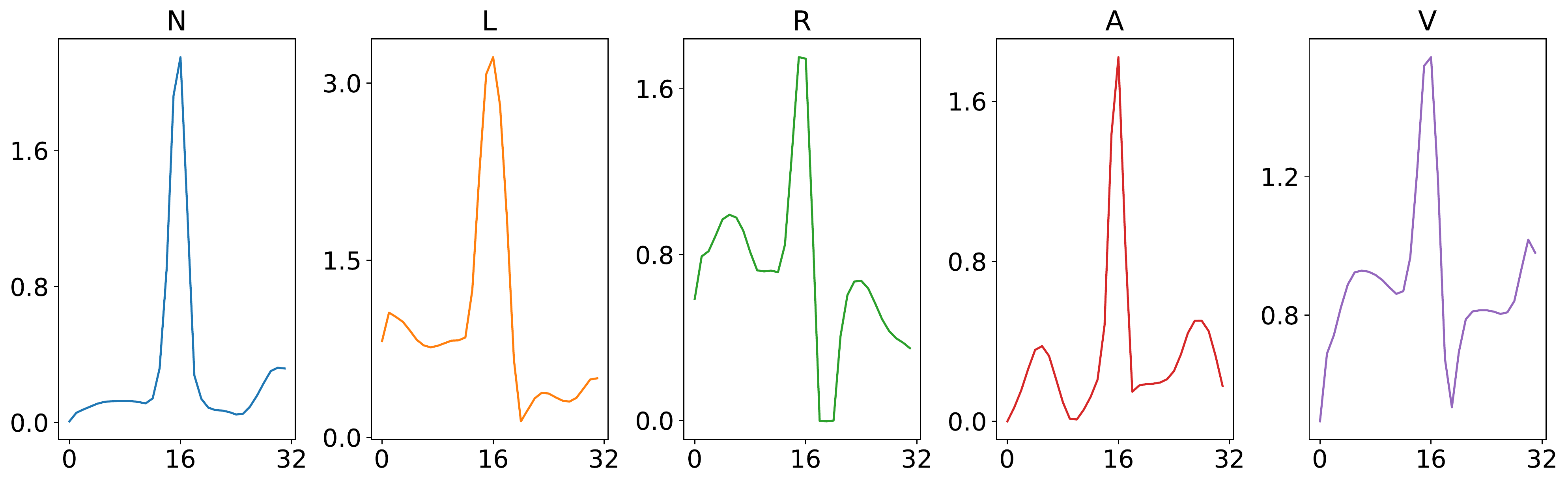}}
	\caption{Visual invertibility. Top row shows raw input data. Bottom row shows one of the channels in the output from the second convolutional layer.}
	\label{fig:visual_invert}
\end{figure}

We first visualize each channel of the split layer output as an initial assessment to observe the possibility of reconstructing the original ECG signals. The model batch size is 32 (i.e., the number of ECG samples fed to the model) and uses 16 filters to extract features. These filters produce the layer activation, which is passed to the server from any split layer.  Fig.~\ref{fig:visual_invert} shows 5 different classes of original ECG samples on the top row vs. the reconstructed version from 5 corresponding filters activation after two layers on the bottom row. We can observe that there is high similarity between them, which means the possibility of significant leakage. 

Our framework goes further to generalise this observation by measuring the correlations between the split layer activation and original samples. To quantify our results, we employ two other metrics: distance correlation and Dynamic Time Warping as explained below.

\subsection{Distance Correlation}\label{sec:4.3}
In statistics, distance correlation is a measure of dependence between two paired vectors of arbitrary dimensions.  It is on the scale of 0-to-1. Distance correlation of 0 refers to independent vectors, whereas 1 means highly dependent and fully similar. Distance correlation has already been used in the context of deep learning to measure the autoencoder correlation \cite{wang2018distance} and reconstruction of raw data from intermediate layers of 2D CNN \cite{vepakomma2019lekage}. Therefore, we apply distance correlation as a measure to monitor the dependency between  each channel of split layer output and corresponding raw ECG signal.  

The distance correlation of two random variables is assessed by dividing their distance covariance by the product of their distance standard deviations, given below \eqref{eq:dcor}~\cite{gretton2009discussion}. 

\begin{equation}\label{eq:dcor}
    \textup{dCor}(X,Y)=\frac{\textup{dCov}(X,Y)}{\sqrt{\textup{dVar}(X)\;\textup{dVar}(Y)}}
\end{equation}

To get the distance correlation between raw input and the split layer activation, we firstly pick 10K of ECG signals from the dataset. Then we take the average of distance correlations from 10K samples, which come between raw ECG signal and split layer output channels. We call this, {\it distance correlation mean}. Because the distance correlation measurement requires two vectors with same dimension, we apply average downsampling on the original ECG signal, to adapt its size to corresponding split layer output.

\begin{figure}[h]
	\centerline
	{\includegraphics[scale=0.36]{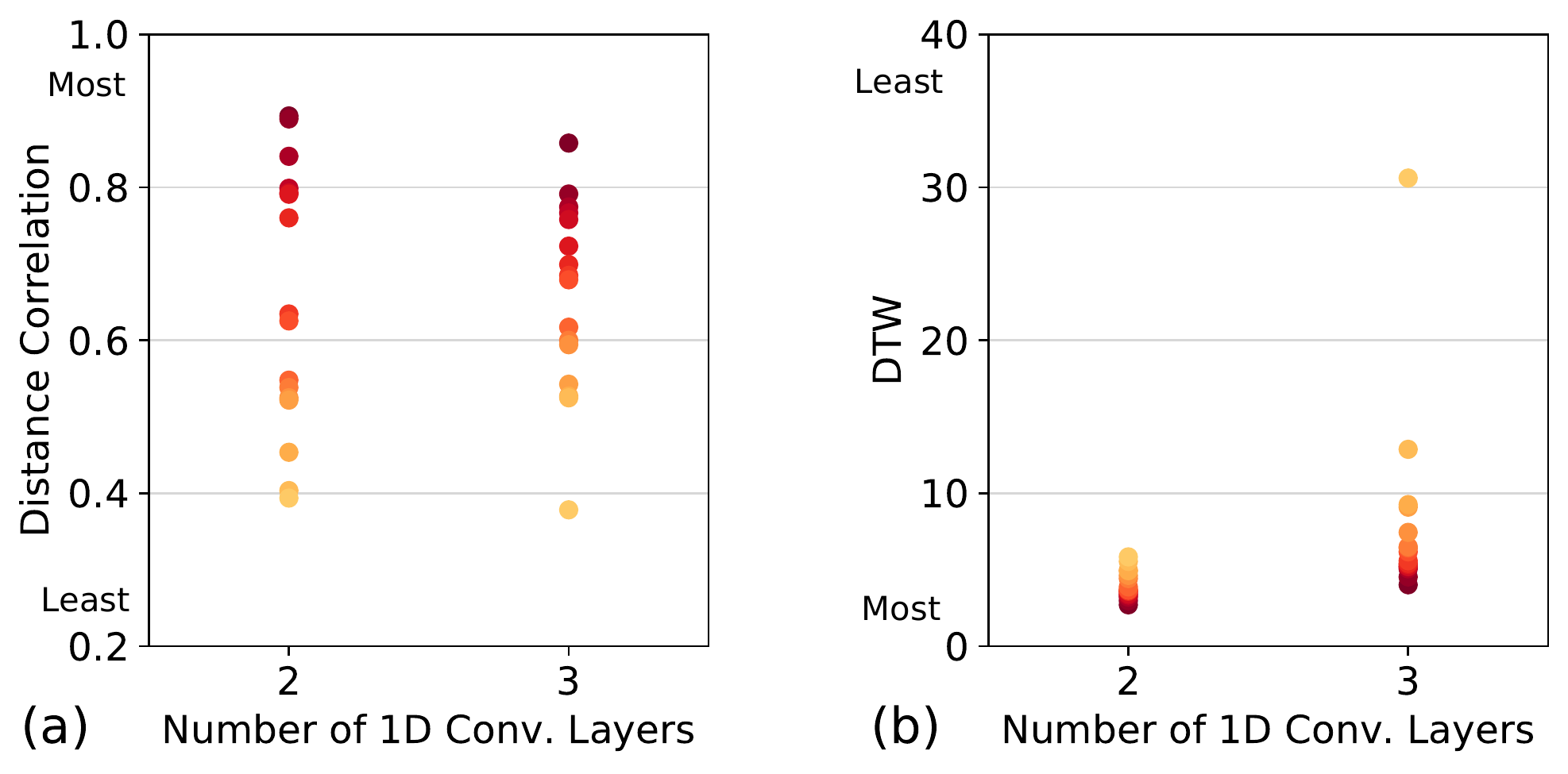}}
	\caption{Distance correlation and DTW between raw input and the activated outputs after two and three 1D convolutional split layers, respectively. Each dot represents the channel of the split layer output.}
	\label{fig:dcor_dtw_leakage}
\end{figure}

Fig. \ref{fig:dcor_dtw_leakage} (a) shows the distance correlation mean between 16 channels of the split layer activation and corresponding raw data. We repeat these experiments after two and three convolutional layers, respectively. Splitting after two layers, the results suggest that the correlation of the top channels activation is very high (0.89), which indicates high leakage and can be exploited to reconstruct the raw data, as shown previously in Fig. \ref{fig:visual_invert}, visual invertibility. Similarly, some channels of the activation after three layers exhibits high dependency, as well. However, the correlation of the top channels is reduced by (0.03); from (0.89) to (0.86) between splitting after 2 and 3 layers. This gives us intuition to increase the number of layers as a mitigation strategy investigated in the next section.

\subsection{Dynamic Time Warping (DTW)}
To generalise our observations, we utilize another well-known similarity measurement in time series analysis called Dynamic Time Warping (DTW) \cite{senin2008dtw}. DTW is an algorithm that can accurately measure the similarity between two temporal sequences, which may vary in speed. It is widely used in speech recognition and signature recognition.

Given two time series $X = (x_1, x_2, ..., x_N)$ and $Y = (y_1, y_2, ..., y_M)$, represented by a sequence of values, DTW algorithm starts by constructing the distance matrix $C \in \mathbb{R}^{N\times M}$ representing all pairwise distances between $X$ and $Y$. This distance matrix, also called as local cost matrix $C_l$, for the alignment of two sequences $X$ and $Y$ is calculated as follows:
\begin{equation}
 C_l \in \mathbb{R}^{N \times M} : c_{ij} = \left \| x_i - y_j \right \|, \; i \in \{1\isep N\}, \; j \in \{1\isep M\} 
\end{equation}

Once the local cost matrix is constructed, the algorithm explores the alignment path (or warping path), which runs through the low-cost areas. The warping path which has the lowest cost associated with alignment is called the optimal warping path. The length of optimal warping path is finally used as a measurement of the similarity between $X$ and $Y$.  Zero-length optimal warping path refers to high similarity, whereas increasing length of optimal warping path toward, e.g., 1000 means  higher dissimilarity. In this paper, we apply DTW as a measure to monitor the similarity between each split layer filter activation and corresponding raw ECG signal. 

For the measurement, we go through the same process mentioned in Section \ref{sec:4.3} but with DTW, and we call this, {\it DTW mean}. Unlike distance correlation, DTW can be computed although two vectors have different sizes, so we do not apply any downsampling on the original data in this case.

Fig. \ref{fig:dcor_dtw_leakage} (b) shows the mean of DTW between the intermediate split layer of all 16 channels activation and corresponding raw data. We also conduct this experiment with split layer output at second and third convolutional layers, respectively. Splitting after two layers, DTW indicates that the similarity of the top channels activation, which also exhibits high leakage and can be exploited to reconstruct the raw data as visually shown previously in Fig \ref{fig:visual_invert}.
Similarly, the channels output after three layers shows high similarity; Again, the top DTW mean increases by 0.21; from 2.70 to 2.98 between splitting after two and three layers. Furthermore, the lowest DTW mean increases significantly to more than 30, by adding one more convolutional layer. This suggests  that increasing the number of layers may reduce the leakage.

\vspace{0.2cm}
\noindent{\bf Summary:} Our  leakage analysis framework to test our RQ 2 via three empirical (visual invertibility) and numerical metrics (distance correlation and DTW)  indicates that  activated output after two and three convolutional layers can be used to reconstruct the raw data. In other words, sharing the  intermediate activation from these layers may result in severe privacy leakage. Therefore, RQ 2 is answered unfavourably. 

\begin{figure*}[htbp]
	\centerline
	{\includegraphics[scale=0.4]{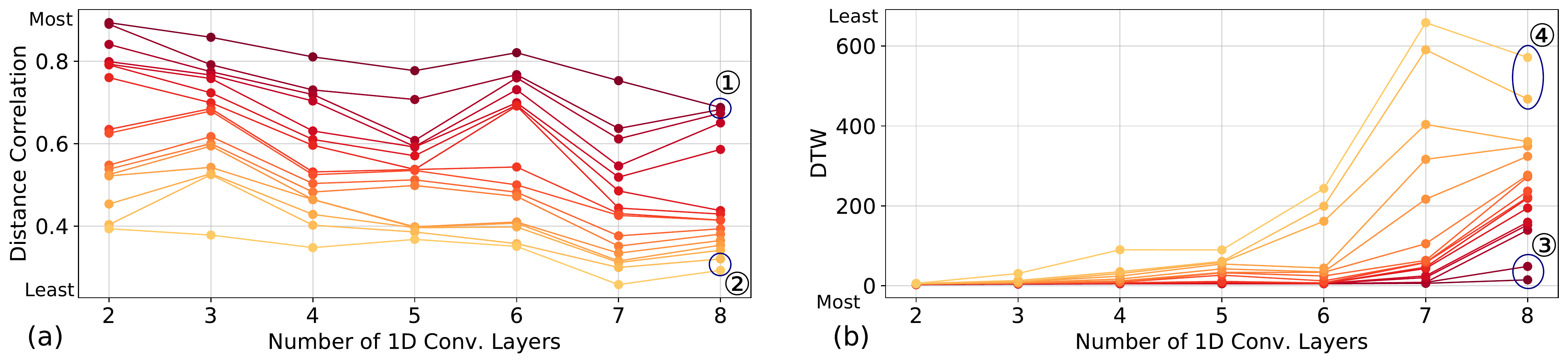}}
	\caption{Mean of distance correlation and DTW calculated for each channel of the split layer output. Each dot represents the channel of the split layer output. Thicker the color of dot, higher the similarity between the channel of the split layer output and corresponding raw data.}
	\label{fig:dtw_dcor_mean}
\end{figure*}

\section{Mitigate the shortcoming? }\label{sec:mitigation}
To further answer our RQ 2, we investigate a number of strategies which can be deployed in 1D CNN model to mitigate the privacy leakage. Specifically, we apply and evaluate two mitigation techniques to reduce potential privacy leakage by i) adding more hidden layers before splitting and ii) applying differential privacy on split layer activation before transmitting them to the server. We measure the efficacy of mitigation techniques via both i) privacy leakage reduction using distance correlation and DTW as well as ii) model accuracy after applying mitigation techniques.

\begin{figure}[!h]
	\centerline
	{\includegraphics[scale=0.45]{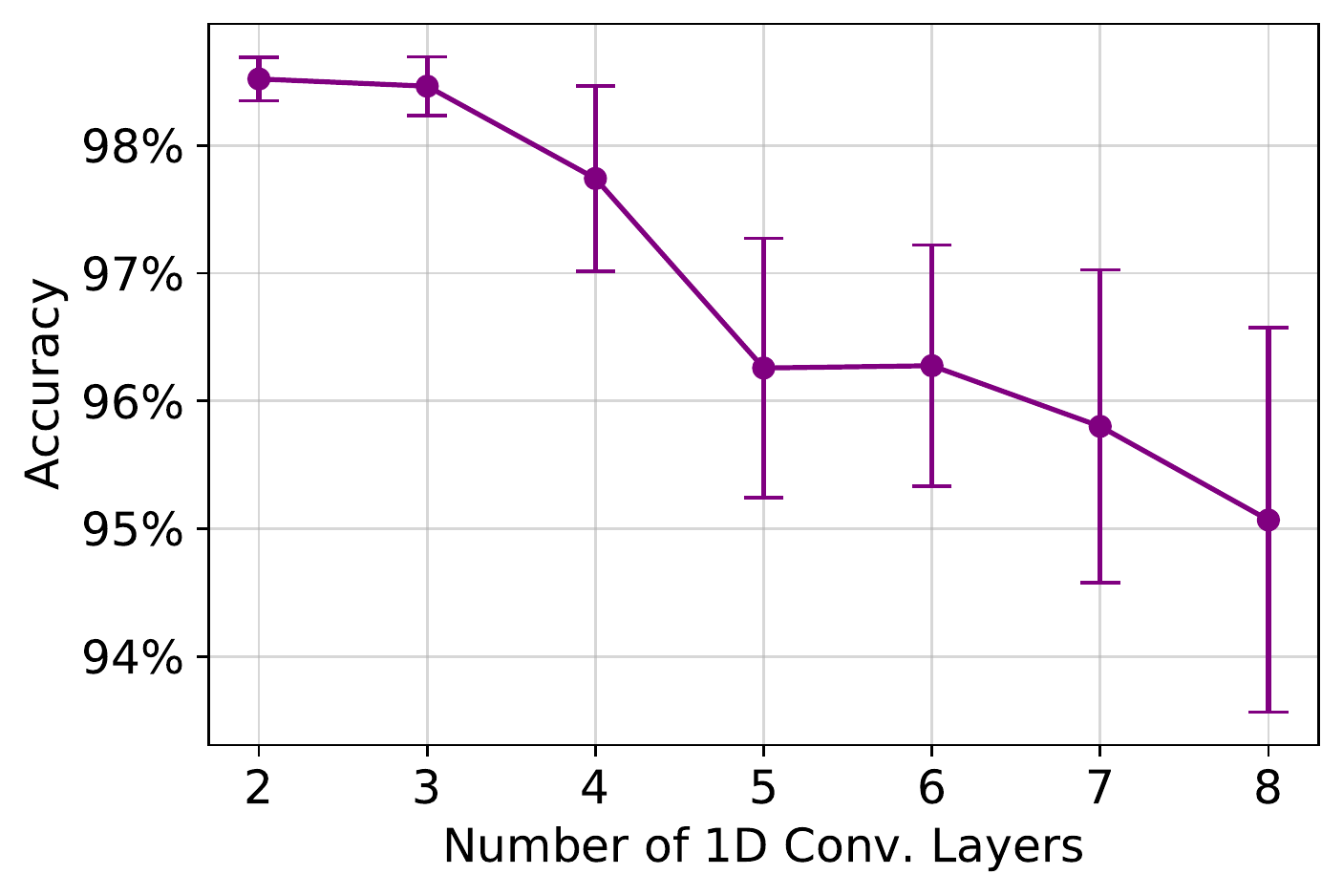}}
	\caption{Mean and standard deviation of test accuracy as incrementing the number of convolutional layers from 2 to 8 for split learning.}
	\label{fig:acc_adding_layer}
\end{figure}

\subsection{Adding More Hidden Layers}
Specifically, we add more convolutional layers---ranging from two to eight---to the client before the split layer. In other words, the model architecture becomes more complex, given the number of layers held by the server being constant. To be consistent, the layers to be added use same configuration as illustrated in \ref{sec3.1.2}.
Each additional convolutional layer utilizes 16 filters whose size is 5. Zero padding is applied to keep the size of output as a power of 2.  Moreover, the size of the filter used for each newly added hidden convolutional layer is 5. Leaky ReLU is also selected for newly added layers. We select activation function as Leaky ReLU, rather than ReLU, to prevent the dying ReLU problem.

\textbf{Distance Correlation Mean:} Fig. \ref{fig:dtw_dcor_mean} (a) shows the mean of distance correlation between the intermediate split layer of all 16 channels output and corresponding raw data after second to eighth convolutional layer. The correlation of each  channel is measured against the corresponding raw ECG signal and represented as a dot in the figure. We then sort the distance correlation mean in descending order on the Y-axis, where distance correlation of 1 indicates a high risk of leakage, and 0 means low risk. It is clear from the top correlated filters that there is a slight reduction in the distance correlation from (0.89) to (0.69)  as the number of hidden convolutional layers increases; however, there is still  some highly correlated channels whose distance correlation means are above 0.5.

\textbf{DTW Mean:} Fig. \ref{fig:dtw_dcor_mean} (b) further shows the mean of DTW between the intermediate split layer of all 16 channels activation and corresponding raw data after second to the eighth layer. The similarity of each filter is calculated against the corresponding raw data ECG and represented as a dot in the figure. We again grade the mean similarity in ascending order from 0 as a high risk of leakage to 600+ as low risk. It is clear from the channels that there is a significant dissimilarity improvement from  zero(0) to (600+) of some channels; However, DTW means of many other  channels are still close to zero, which indicate high leakage and can be potentially exploited to reconstruct the raw ECG data.

\begin{figure}[!h]
	\centerline
	{\includegraphics[scale=0.45]{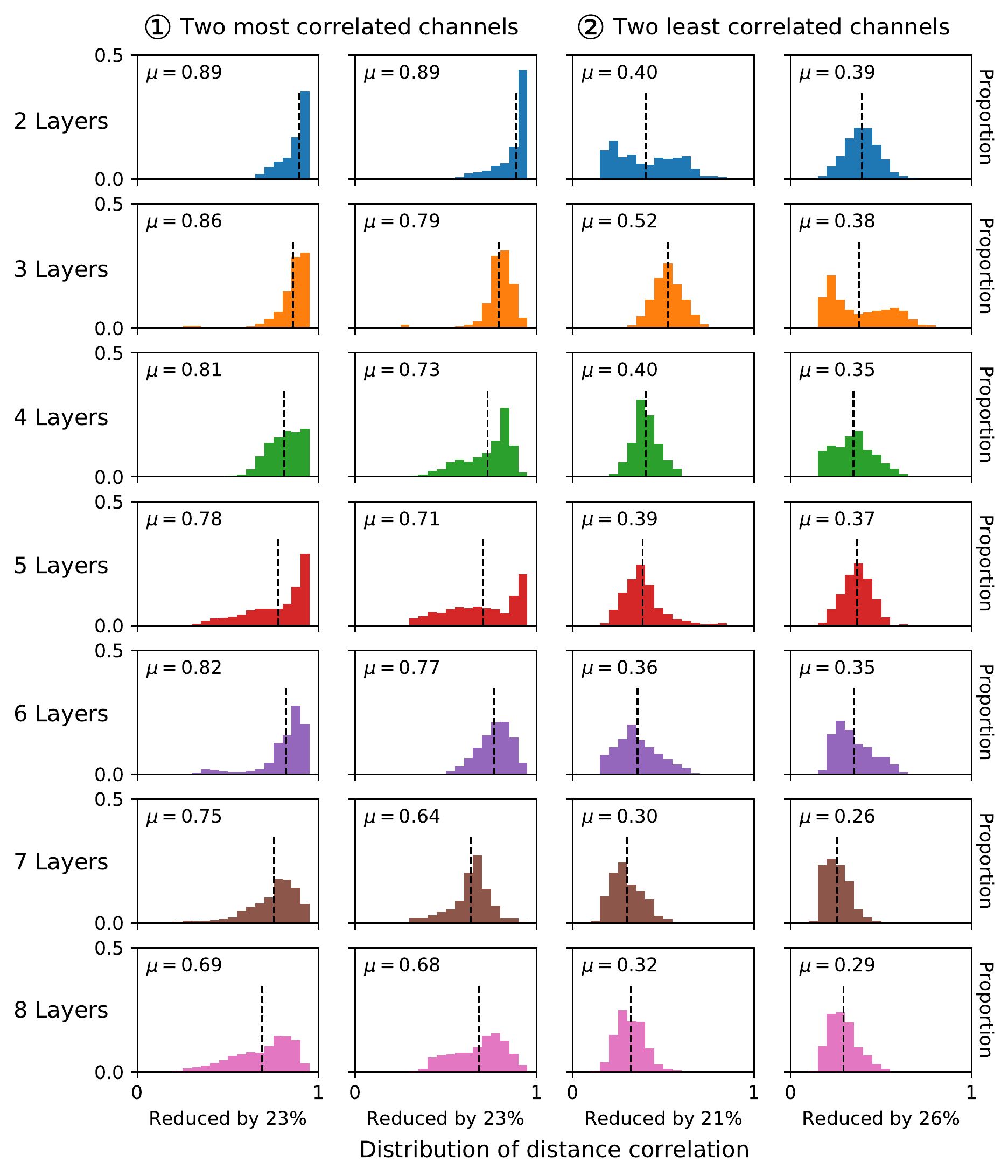}}
	\caption{Effects of adding more layers on distance correlation distribution of the two most correlated channels, \textcircled{\raisebox{-0.6pt}{1}} in Fig.~\ref{fig:dtw_dcor_mean}, and two least correlated channels, \textcircled{\raisebox{-0.6pt}{2}} in Fig.~\ref{fig:dtw_dcor_mean}. For each distribution, $\mu$ denotes the mean. Privacy leakage mitigation by increasing the number of layers seems effective for least correlated filters, but it is ineffective for those showing high distance correlation, whose correlation is still high, more than 0.68, when 8 convolutional layers run on the client.}
	\label{fig:dcor_shift}
\end{figure}

\begin{figure}[!h]
	\centerline
	{\includegraphics[scale=0.4]{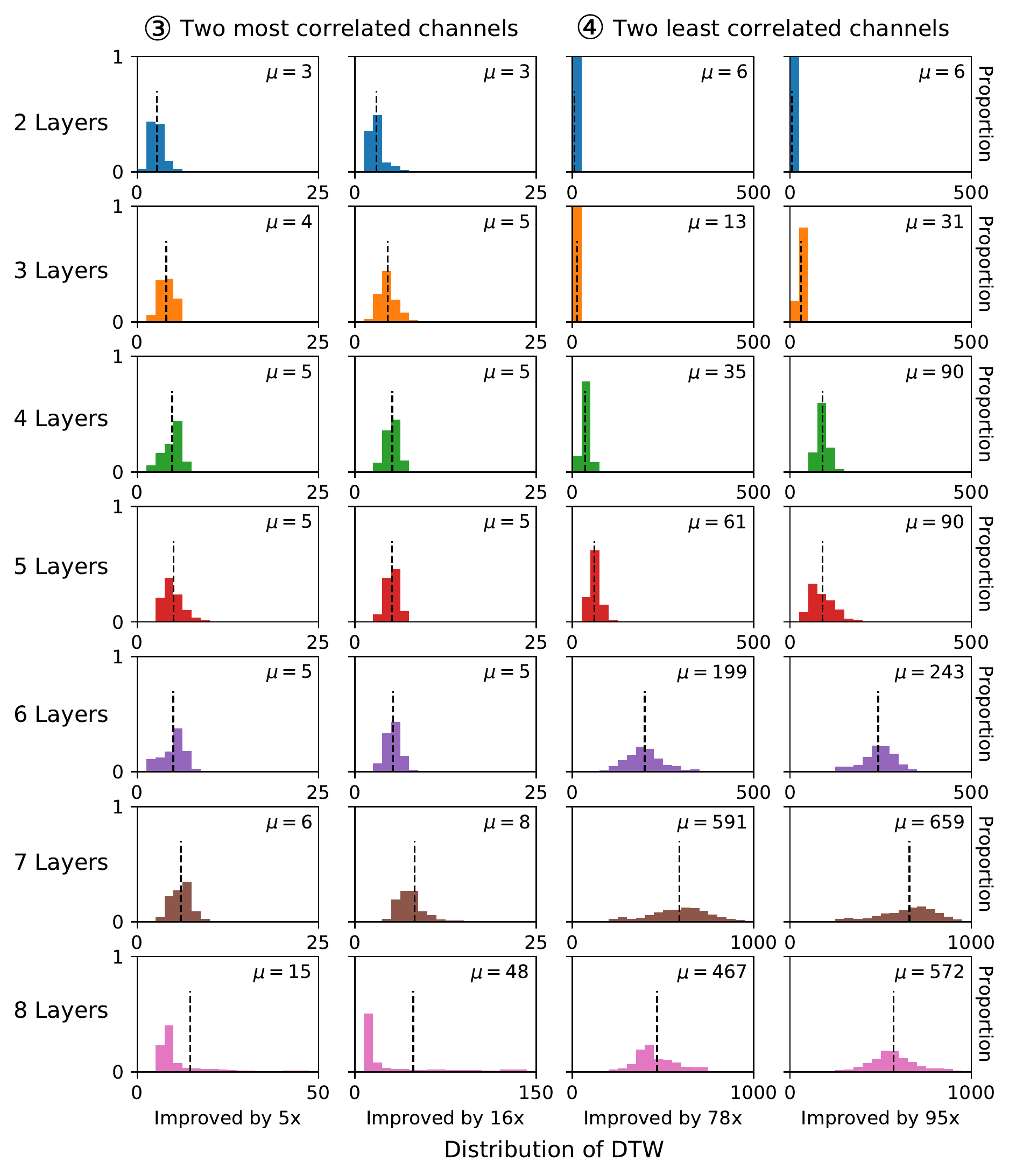}}
	\caption{Effects of adding more layers on DTW distribution of the two most correlated channels, \textcircled{\raisebox{-0.6pt}{3}} in Fig.~\ref{fig:dtw_dcor_mean}, and two least correlated channels, \textcircled{\raisebox{-0.6pt}{4}} in Fig.~\ref{fig:dtw_dcor_mean}. For each distribution, $\mu$ denotes the mean. Privacy leakage mitigation by increasing the number of layers seems effective for least correlated filters---DTW improved by e.g., 84 times, but it is ineffective for those showing low DTW, whose improvement is less by, e.g., 5 times, when 8 convolutional layers run on the client.}
	\label{fig:dtw_shift}
\end{figure}

\textbf{Distribution:} We further investigate the distribution of distance correlation, as detailed in Fig. \ref{fig:dcor_shift}, which illustrates two most correlated  channels vs. two least correlated channels. The distance correlation distribution is continuously reduced but seems ineffective to protect the highly correlated channels. Similarly, Fig. \ref{fig:dtw_shift} presents DTW distribution of the most two correlated channels vs. the least two correlated channels. The least correlated shows clear improvements by e.g., 84 times; however, DTW also emphasizes that increasing the number of layers seems ineffective with the most highly correlated channels i.e., improved only by 5 times. 

\begin{figure*}[!t]
	\centerline
	{\includegraphics[scale=0.4]{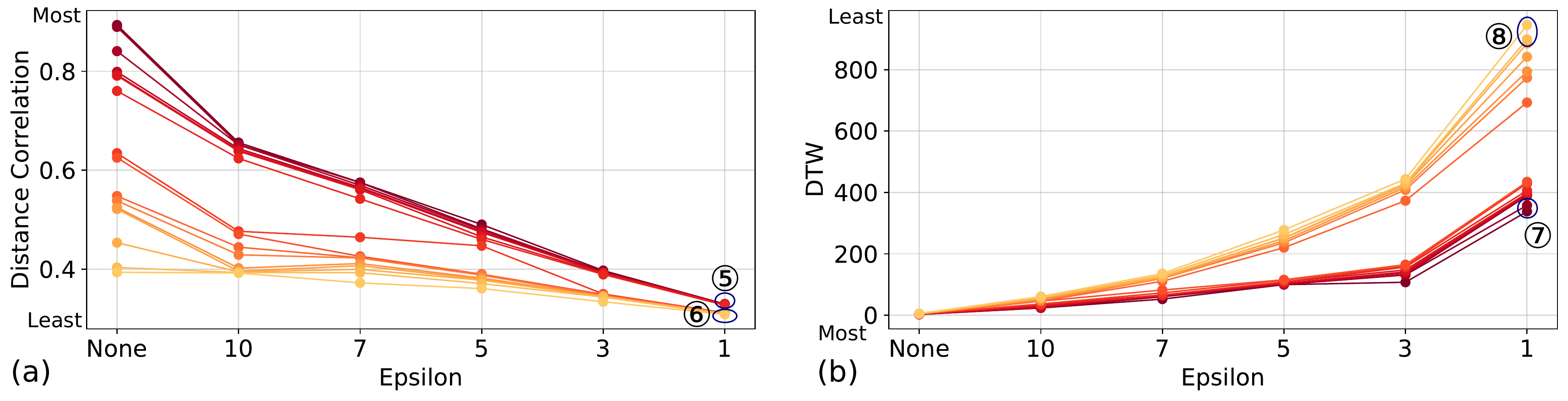}}
	\caption{Mean of distance correlation and DTW after applying differential privacy on the split layer output before transmitting the activation. Each dot represents the channel of split layer output. Thicker the color of dot, higher the similarity between the channel of the split layer output and corresponding raw data.}
	\label{fig:dtw_dcor_mean_diffpriv}
\end{figure*}

\subsection{Applying Differential Privacy on Split Layer}
As the second mitigation technique, we apply differential privacy on split layer filters activation before transmitting them to the server. Differential privacy has already been widely used in deep learning to protect privacy \cite{abadi2016deep}. Given input space $X$, output space $Y$, privacy parameter $\epsilon$, and a randomisation mechanism $\mathcal{M}$. We say $\mathcal{M} : X \rightarrow Y$ is $\epsilon$-differentially private if, for all neighbouring inputs $X\simeq X^{'}$ and all sets of outputs $S \subseteq Y$ satisfy the following:
\begin{equation}
   \Pr\left [ \mathcal{M}(X) \in S \right ] \leq e^\epsilon \times \Pr\left [ \mathcal{M}(X') \in S \right ]
\end{equation}
Therefore, the value of $\epsilon$ determines the strength of privacy. Specifically, we employ the Laplace differential privacy mechanism on the split layer activation, which is widely used for numerical data \cite{holohan2019diffprivlib}. It adds noise from the Laplace distribution, which can be expressed by a probability density function. The level of noise relies on pre-determined $\epsilon$ on a scale of 10 (weakest privacy)-to-0 (strongest privacy where the data cannot be used). 

\begin{figure}[h]
	\centerline
	{\includegraphics[scale=0.4]{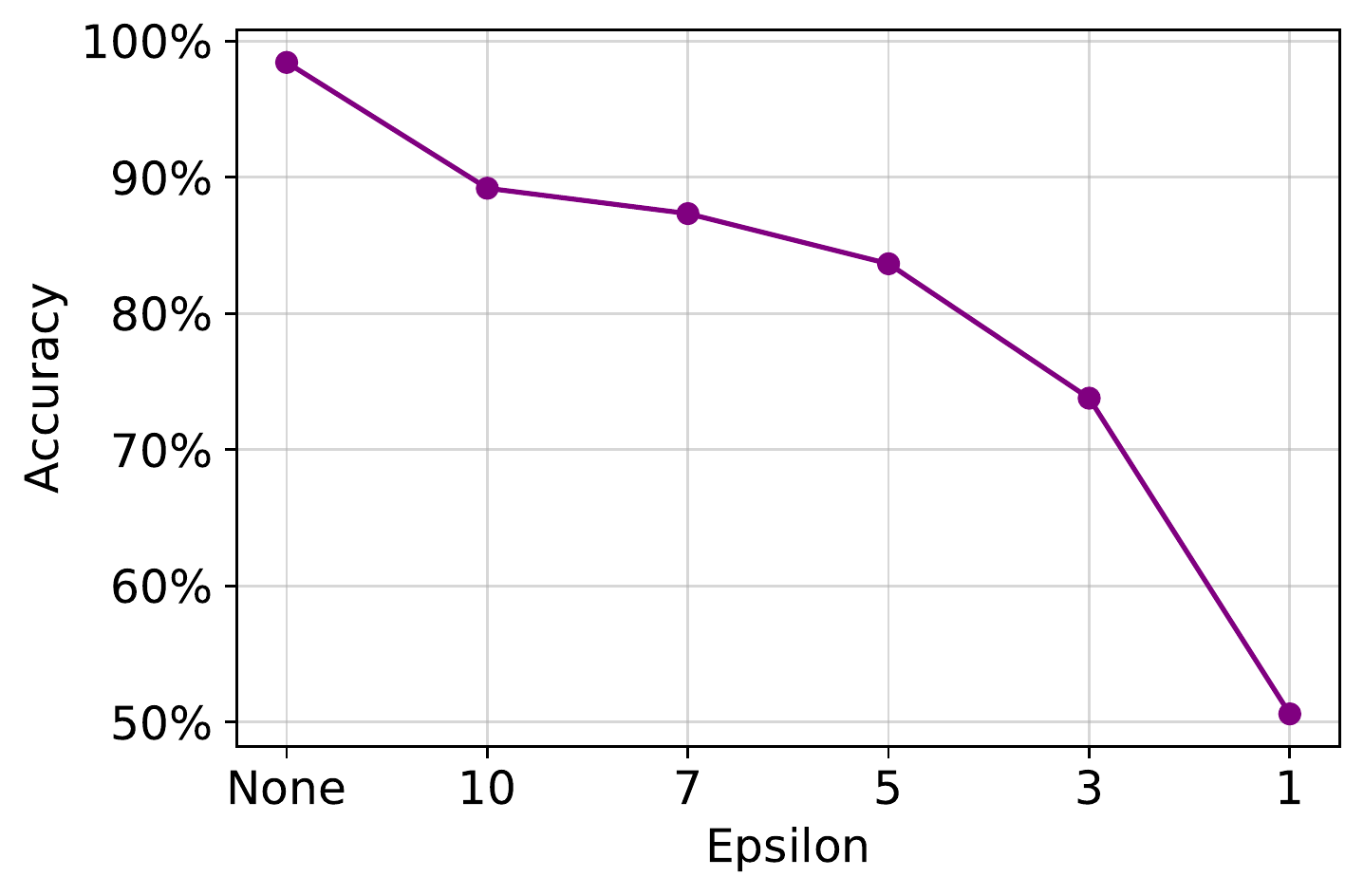}}
	\caption{Accuracy changes after applying differential privacy with stronger epsilon values between 10 (weakest) and 1 (strongest).}
	\label{fig:accuracy_diffpriv}
\end{figure}

\textbf{Distance Correlation Mean:} Fig. \ref{fig:dtw_dcor_mean_diffpriv} (a) shows the mean of distance correlation between the intermediate split layer of all 16 channels activation  after applying different $\epsilon$ levels of differential privacy, and corresponding raw data. It is under the expectation that the strongest differential privacy level of, e.g., $\epsilon=1$ can fully protect all the output channels; However, it comes with the cost of degrading the classification accuracy significantly from 98.9\% to only 50\%  as shown in Fig.\ref{fig:accuracy_diffpriv}.


\textbf{DTW Mean:}  Fig. \ref{fig:dtw_dcor_mean_diffpriv} (b) shows the mean of DTW between the intermediate split layer of all 16 channels activation  after applying various $\epsilon$ levels of differential privacy, and corresponding raw data. The results also suggest that strongest differential privacy level of $\epsilon=1$ can increase the dissimilarity between the filter activation and corresponding raw data which protects them against potential reconstruction attempts; However, strongest privacy level of $\epsilon=1$ damages the accuracy significantly as explained earlier.

\begin{figure}[h]
	\centerline
	{\includegraphics[scale=0.45]{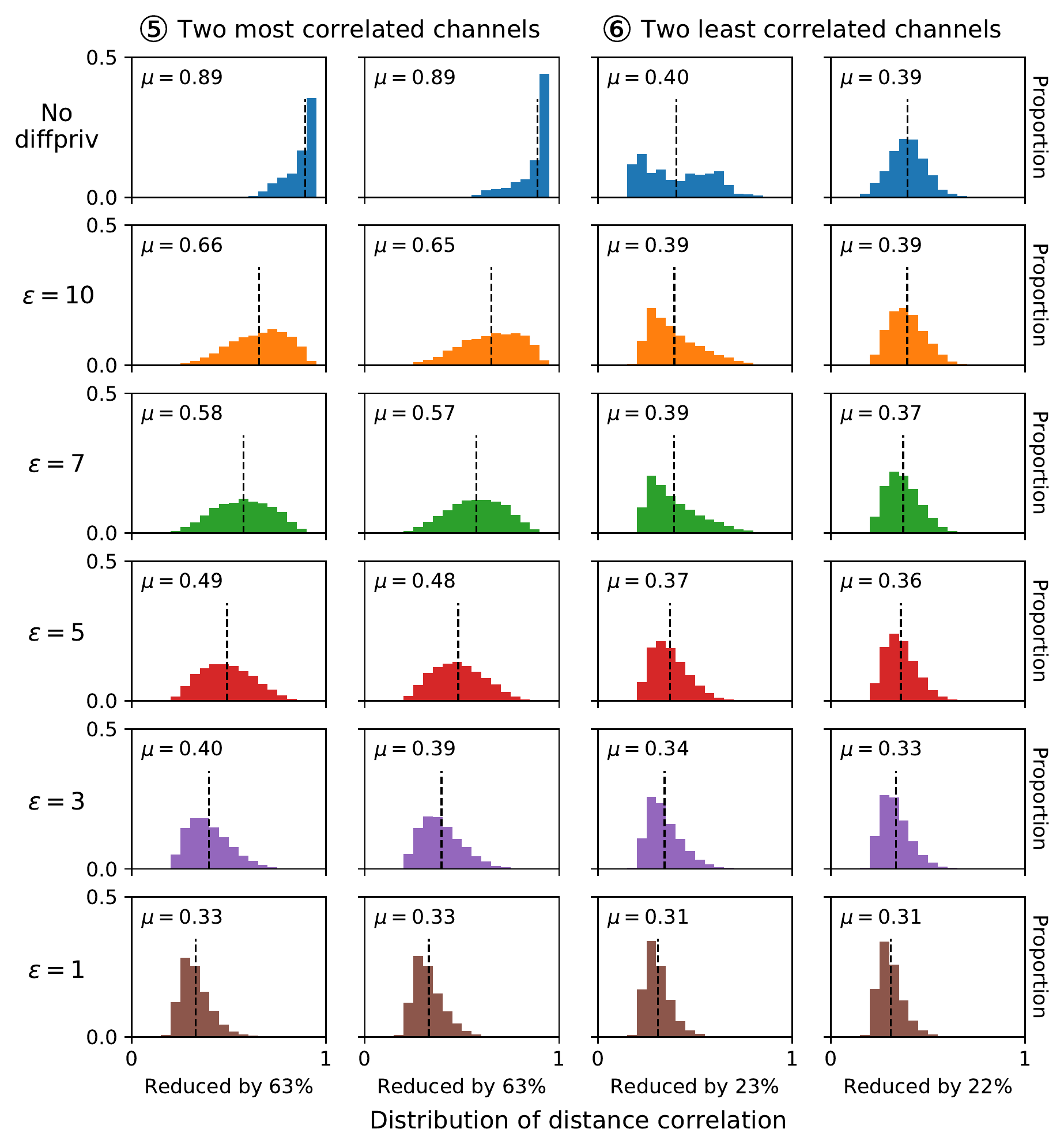}}
	\caption{Effects of applying differential privacy on distance correlation distribution of two most correlated channels, \textcircled{\raisebox{-0.6pt}{5}} in Fig.~\ref{fig:dtw_dcor_mean_diffpriv}, and two least correlated channels, \textcircled{\raisebox{-0.6pt}{6}} in Fig.~\ref{fig:dtw_dcor_mean_diffpriv}. For each distribution, $\mu$ denotes the mean. In contrast to applying more hidden layers, privacy mitigation by applying differential privacy seems effective for highly correlated channels---reduced by 63\%, but it is ineffective for channels having lower distance correlation, e.g., about 23\%.}
	\label{fig:dcor_shift_diffpriv}
\end{figure}

\begin{figure}[h]
	\centerline
	{\includegraphics[scale=0.45]{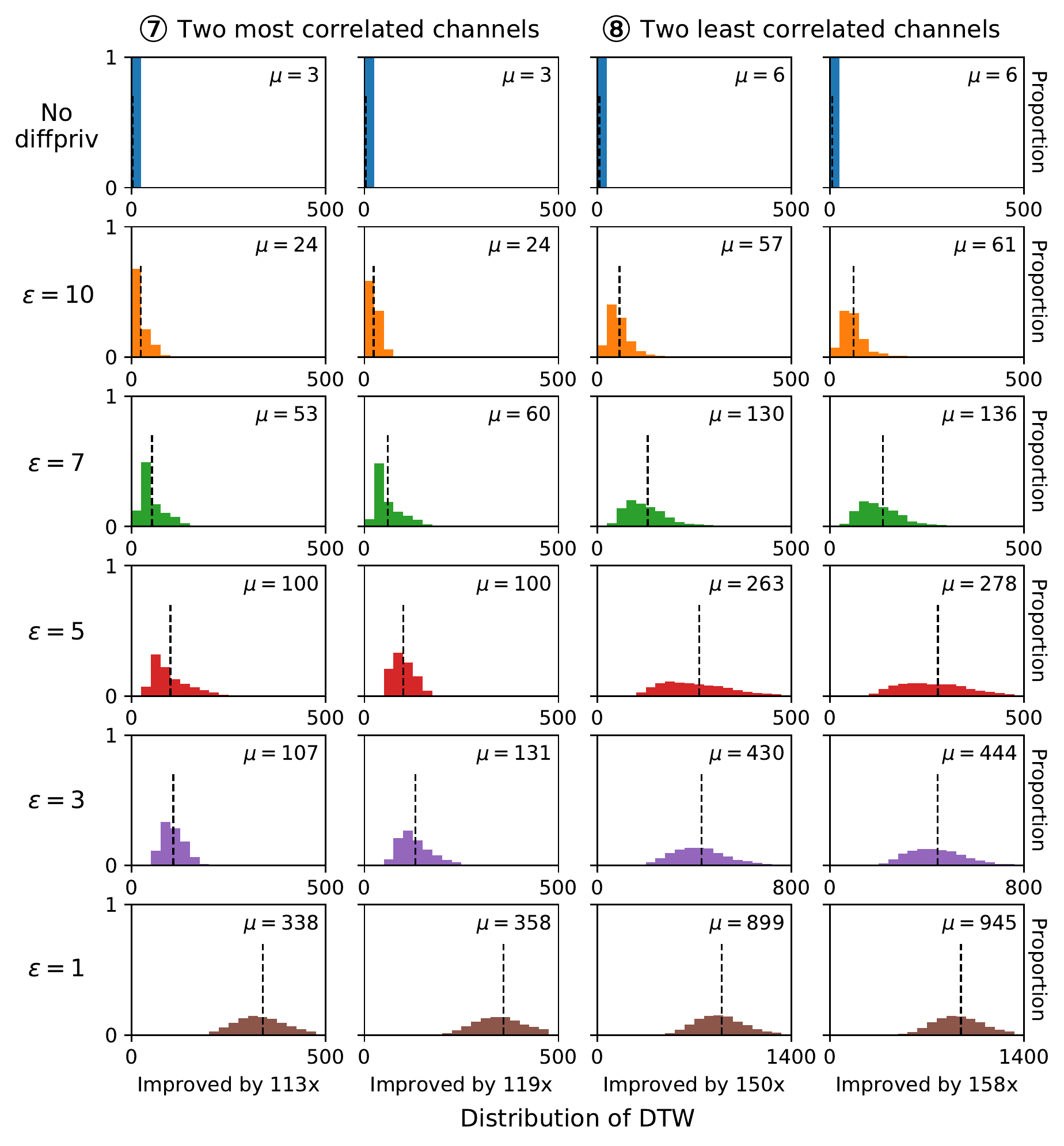}}
	\caption{Effects of applying differential privacy on DTW distribution of two most correlated channels, \textcircled{\raisebox{-0.6pt}{7}} in Fig.~\ref{fig:dtw_dcor_mean_diffpriv}, and two least correlated channels, \textcircled{\raisebox{-0.6pt}{8}} in Fig.~\ref{fig:dtw_dcor_mean_diffpriv}. For each distribution, $\mu$ denotes the mean. In contrast to applying more hidden layers, privacy mitigation by applying differential privacy seems effective for channels having low similarity---improved 158 times, but it is less effective for highly correlated channels, e.g., about 113 times.}
	\label{fig:dtw_shift_diffpriv}
\end{figure}

\textbf{Distribution:} To look closely at the impact on filters after various $\epsilon$ levels, we also deep dive into the distribution of distance correlation and DTW. Fig. \ref{fig:dcor_shift_diffpriv} presents the distance correlation of two most correlated channels vs. two least correlated channels  without differential privacy, using two-layer model. The distance correlation distribution is continuously improved after applying stronger $\epsilon$ (from 10, 7, 5, 3, to 1) on activation before transmitting. Also, Fig. \ref{fig:dtw_shift_diffpriv} shows DTW of two most correlated channels vs. two least correlated  channels. The DTW also improved after applying stronger $\epsilon$ noise; However, it seems less effective for highly correlated channels.

\vspace{0.2cm}
\noindent{\bf Summary:}  None of the two applied mitigation techniques can efficiently mitigate the privacy leakage from all  channels of the split layer activation---it appears that the mitigation is dependent on the applied mitigation techniques as well as convolution filters. On top of that, both of them come with a cost of accuracy reduction of the joint model, which appears not acceptable especially in the differential privacy case. Therefore, the RQ 2 is still hard to be answered favourably even when our mitigation attempts are applied.

\section{Discussion and Future Work}\label{sec:discussion}
\vspace{0.2cm}
\noindent{\bf Why split learning is ineffective to preserve the privacy for 1D CNN?}
To reduce the privacy leakage from the split layer revealed by us, we have consequently applied two specific techniques: i) increasing the number of convolutional layers at the client-side, and ii) applying differential privacy to the split layer output before transmission. However, both techniques suffer the reduction of model accuracy, especially with differential privacy. For the first technique of increasing the number of convolutional layers to the client, it eventually renders the other trade-off that is the computational overhead of the client. As a matter of fact, saving the computational overhead of the client is the other main motivation of split learning since the client only needs to train a small fraction of parameters of the whole model in comparison with the federated learning required to train the whole model~\cite{gupta2018distributed}. Therefore, increasing the number of layers before the split layers to the client further diminishes the benefits of the split learning from the client computational overhead perspective, especially, when the split learning is mounted on resource-restricted devices/agents.

\vspace{0.2cm}
\noindent{\bf Framework to evaluate split learning models:}
We use three metrics to measure the privacy leakage and explore the possibility of reconstructing the ECG samples' raw data from split layer activation. These measurements are visual invertibility, distance correlation, and Dynamic Time Warping. The first measurement is to show an empirical way to reconstruct the raw data from activation. The second and third are used for quantification. We believe these three metrics can be used as a general framework to evaluate the potential privacy leakage from split learning models.
 
\vspace{0.2cm}
\noindent{\bf How to develop more effective split learning for 1D CNN?}
This is an open question that is remaining to be answered as we do not have an affirmative answer. However, we suggest two, but not limited to,  future strategies. (i) Our first mitigation attempt to increase the number of hidden layers shows that it is possible to protect many of the channels in the activated output, but not all of them without significant accuracy degradation. One would explore leakage mitigation techniques for only a few revealing output channels. (ii) Existing split learning follows a vertical split mechanism, which means all the split layer output should be shared. To reduce the exposure of all activation, one would explore a multiple-horizontal split and share only the protected channels with the server.

\vspace{0.2cm}
\noindent{\bf Can split learning be applied to LSTM and/or RNN?} As a matter of fact, employing LSTM~\cite{hochreiter1997long} and/or RNN~\cite{mikolov2010recurrent} served as the first trial when we intended to investigate the practicality of dealing with sequential/time-series data. However, we realized that LSTM and RNN are sequential networks, while the current split technique is always vertically split. Therefore, we found that there is no efficient means of splitting LSTM and RNN. This eventually motivated us to  adopt 1D CNN that can be vertically split as well to deal with pervasive sequential/time-series data. Whether split learning can be properly applied to LSTM and/or RNN remains to be further investigated.

\vspace{0.2cm}
\noindent{\bf Limitations:}
Our work has two main limitations. Firstly, we explore the leakage in 1D CNN using one sensitive health application, which is ECG biomedical signals. We experiment with the ECG heartbeat samples from the widely-used MIT-BIH dataset.  Other 1D CNN applications and datasets remain to be investigated. Secondly, privacy leakage from the split layer of 2D CNN various applications remains to be addressed.

\section{Related Work}\label{sec:relatedwork}
In~\cite{gupta2018distributed}, the authors envision that it should be challenging to reconstruct the data localized on the client-side when split learning is employed. The main argument, generally, is that the server can not access the weight of split layers held by the client---this is the case for federated learning. Inverting the weight is computationally infeasible in practice. While their analysis focuses on 2D CNN, this paper has shown that reconstructing the raw data in 1D CNN is more than possible.

Vepakomma et al. \cite{vepakomma2019lekage} investigated the potential privacy leakage of split learning in 2D CNN using the MNIST dataset. They also found high potential leakage from the split layer. They showed that by slightly scaling the weights of all layers before the split, it is possible to reduce the distance correlation between the split layer activation and raw data. This scaling mechanism is effective in 2D CNN because of the large number of hidden layers before the split layer. However, existing 1D CNN models such as in \cite{kiranyaz2015real,li2017classification} have only 2-3 hidden layers, which renders scaling their weights ineffective and degrades the accuracy. 

The other privacy leakage that has been revealed is the membership inference. For example, Melis {\it et al.}~\cite{melis2019exploiting} demonstrated the membership inference attack on federated learning by observing the gradients aggregated from the model trained on clients. It is not surprising to envision such a membership inference attack could be applicable to split learning. To which extent the split learning can be resistant to inference attack leaves interesting future work.

Other than privacy concerns of distributed learning, there are also emerging security concerns inherent to the distributed learning, in particular, the backdoor attacks~\cite{gao2019strip,wang2019neural,gao2019design}. Generally, a backdoor attack occurs when the training data is tampered by a malicious party under the model training outsource scenario. The backdoor model behaves normally for clean inputs while misclassifying any input to the target class when the input is stamped with a trigger. It has been shown that federated learning is inherently vulnerable to such a backdoor attack because the server has no control over the local data by design~\cite{bagdasaryan2018backdoor,sun2019can}. Therefore, participants can manipulate their data as they want to insert a backdoor to the joint model. We believe that split learning, for the same reason, is also inevitably vulnerable to backdoor attacks from the security perspective besides the privacy leakage revealed in this work. Thus, it is important to consider this security concern when deploying split learning in realistic security-critical applications.

\section{Conclusion}\label{sec:conclusion}
In this paper, we explored the feasibility of split learning to deal with sensitive time-series data in particular personal ECG signals to detect heart abnormalities. We introduced the first implementation of split learning into the 1D CNN model. We proposed a privacy assessment framework for CNN models when using split learning, with three metrics: visual invertibility, distance correlation, and DTW. Based on this framework, we extensively evaluated the privacy leakage exemplified by the ECG dataset. Our results demonstrate that adopting split learning directly into 1D CNN models for time-series/sequential data would exhibit a possibility of high privacy leakage from feature values. Initial mitigation attempts via i) increasing the number of layers in a CNN model and ii) using differential privacy explicitly indicate there is a trade-off between the degree of privacy leakage reduction and joint model accuracy deterioration---substantial when applying the differential privacy. Perhaps, it would be more challenging to preserve privacy via 1D CNN models compared with 2D CNN models. This is because 1D CNN usually has much less hidden layers where more 1D CNN layers usually tend to affect the model accuracy adversely. As future work, privacy leakage through the split layer should be thoroughly evaluated for (1D and 2D) CNN models, and corresponding effective mitigation techniques need to be developed.

\section*{Acknowledgment}
The work has been supported by the Cyber Security Research Centre Limited whose activities are partially funded by the Australian Government’s Cooperative Research Centres Programme. This work was also supported in part by the ITRC support program (IITP-2019-2015-0-00403). The authors would like to thank all the anonymous reviewers for their valuable feedback.

\bibliographystyle{unsrt}
\bibliography{References}

\appendix

\section{Appendix: ECG Dataset Preprocessing in Detail} \label{sec:appendixA}

The MIT-BIH database contains 48 records retrieved from 47 different patients. About 110K ECG signals are distributed in 48 records. Each record includes a 30-minute excerpt of two-channel ECG signals. Also, each record has an annotation pair that contains time positions and beat types of all ECG signals in the paired record. Similar to \cite{kiranyaz2015real, li2017classification}, we first eliminate four records (record 102, 104, 107, and 217) containing paced heartbeats. Among two channels of ECG data, we extract only the upper channel from each record as it highlights abnormal ECG patterns \cite{li2017classification}. In most cases, the upper channel signals come from modified limb II (ML II); however, in the case of record 114, the ML II signal is located at its bottom channel. Therefore, we exclude record 114 too. A total of 43 out of 48 records are selected for further preprocessing.

Each ECG record contains $n$ number of cardiac cycles of heartbeat collected using electrodes placed on the skin. Each heartbeat has three main components: the P wave that reflects the depolarization of the atria; the QRS complex that reflects the depolarization of the ventricles; and the T wave that reflects the repolarization of the ventricles. Together, they can tell if the heart is normal or has a problem.

The next step is to extract every single beat from selected 43 records. We took an equal number of samples from the left and right side of R-peaks which is done by iterating through the time-series annotations. One hundred values were taken from both sides; However, the segment was discarded if another R-peak existed in the sampled interval. The current beat after this step has 201 sampling values containing only single R-peak. Each signal was then rescaled by the min-max normalization shown in Equation \eqref{eq:minmaxnorm}, where $x$ is original, and $x'$ is normalized value in the signal. The normalized signal was downsampled to 128 by adopting the Fourier method.

\begin{equation}
    \label{eq:minmaxnorm}
    x'=\frac{x-\min (x)}{\max (x) - \min (x)}
\end{equation}
Denoising was the last step of refining ECG signals. We chose biorthogonal wavelet as decomposition and reconstruction wavelet function. The level of signal decomposition was 3. As shown in Equation \eqref{eq:softthresh}, we applied a soft thresholding technique to decomposed coefficients $w$, based on calculated universal threshold~\cite{donoho1994ideal}.
\begin{equation}
    \label{eq:softthresh}
    w'=
    \begin{cases}
    \textup{sgn}(w)(|w|-\lambda) & (|w|\geq\lambda)\\[4pt]
    0 & (|w|<\lambda)
    \end{cases}
\end{equation}
\begin{equation}
    \label{eq:univthresh}
    \lambda=\sigma\sqrt{2\log N}
\end{equation}
Equation \eqref{eq:univthresh} shows the universal threshold value, where $\sigma$ is the median absolute deviation of the wave coefficients at the last level divided by 0.6745, and $N$ is the length of the denoising input signal, in this case, 128.

ECG heartbeat samples in the MIT-BIH database are classified into 17 categories according to morphological patterns in a signal, labeled by independent cardiologists. Following \cite{li2017classification}, we select 5 types of heartbeat as classification targets: $N$ (normal beat), $L$ (left bundle branch block), $R$ (right bundle branch block), $A$ (atrial premature contraction), $V$ (ventricular premature contraction). After excluding beat samples whose labels are not included among 5 types, 96K beat samples remain. We randomly pick 6,000 samples for $N$, $L$, $R$, and $V$ class, however, 2,490 samples for $A$ class, as a number of $A$-labeled beats in the dataset are not as sufficient as other categories. We divide this data pool equally into the train and test set as shown in Table \ref{tb:dataset}.

\section{\textbf{Appendix: Training Flow in Detail.}}\label{sec:appendixB}
We further detail the training flow in Algorithm \ref{algorithm1} and Algorithm \ref{algorithm2} in terms of 1D CNN, as shown in Fig. \ref{fig:flow}. 
This training flow can be directly applied to the 1D CNN ECG classification model we made in Section \ref{sec3.1.2}.

In initialization, the client first connects to the server through the socket and gets some training parameters to synchronize. The initialization in Algorithm \ref{algorithm1} and \ref{algorithm2} require 5 parameters to be synchronized between the client and the server. The server sends $\phi$, $\eta$, $o$ and $n$ to the client that the client should adjust to, just after accepting the connection. $\phi$ is a random weight initializer, $\eta$ is learning rate, $o$ is a type of optimizer, and $n$ is the batch size. These four hyperparameters should be synchronized on both sides to let them trained in which is the same way. When the client receives $n$, it sends $N$ to the server, which is the number of total batches to be trained. $n$ is implicitly used not only for the batch generation but also for determining the shape of the matrix in forward and backward propagation. $N$ helps the server to know how many times the forward  backward propagation should be done. If needed, the number of epochs should also be synchronized between the client and the server. In Algorithm \ref{algorithm1} and \ref{algorithm2}, they give only a single epoch training process. Hence, the training repetition process is omitted.

Forward propagation starts from the client-side. The client first generates the batch, $(x,y)$, which has $n$ data extracted from the train set $\mathbb{D}$. $x$ represents the features of data, and $y$ shows their labels. Starting with $x$, the client does the forward propagation until layer $l$. If layer $i$ is 1D convolutional layer, $f^{(i)}$ can be expressed as follows~\cite{kiranyaz2015real}:
\begin{equation}\label{eq:conv1dforward}
z^{(i)}_k=f^{(i)}(a^{(i-1)})=b^{(i)}_k+\sum_{j=1}^{C_{i-1}}\textup{Conv1D}(a^{(i-1)}_j, w^{(i)}_{jk})
\end{equation}
where $z^{(i)}_k$ is the $k$-th channel of the output from the $i$-th layer, and $a^{(i-1)}_j$ is the $j$-th channel of activation from layer $i-1$. Also, $w^{(i)}_{jk}$ is a convolution filter which connects between $a^{(i-1)}_j$ and $z^{(i)}_k$. $C_{i-1}$ means the number of channels in the output from the $(i-1)$-th layer. \textup{Conv1D} means the regular 1D convolution operation without zero padding. $b^{(i)}_k$ is the bias value added after the convolution operation. Let us assume that the server has at least one convolutional layer on its previous side. Equation \eqref{eq:conv1dforward} tells that forward propagation on layer $i$ only depends on the activated output of $(i-1)$-th layer. In other words, the server can continue forward propagation on layer $l+1$ by only receiving $a^{(l)}$ from the client. After receiving $a^{(l)}$ and label $y$, the server is able to do forward propagation to the $L$-th layer.

Before starting the backpropagation, the server calculates loss between $a^{(L)}$ and $y$. Here, the loss function $\mathcal{L}$ produces $E$ by computing either mean squared error or cross-entropy loss in the usual case. The backpropagation function $f^{(i)}_{\textup{T}}$ can be written as follows if $i$-th layer is 1D convolutional  layer~\cite{kiranyaz2015real}.
\begin{equation}\label{eq:1dconvbackward}
\frac{\partial E}{\partial a^{(i)}_k}=f^{(i)}_{\textup{T}}(\frac{\partial E}{\partial z^{(i+1)}_k})=\sum_{j=1}^{C_{i+1}}\textup{Conv1Dz}(\frac{\partial E}{\partial z^{(i+1)}_k}, \textup{rev}(w_{kj}^{(i+1)}))
\end{equation}
\textup{rev} means reversing 1D convolution filter array, and \textup{Conv1Dz} is full 1D convolution operation with $(\text{filter size} - 1)$ zero pads on both side. Again, Equation \eqref{eq:1dconvbackward} indicates that computing $\frac{\partial E}{\partial a^{(i)}}$ only depends on $\frac{\partial E}{\partial z^{(i+1)}}$. This means the client can continue backpropagation by receiving just $\frac{\partial E}{\partial a^{(l)}}$ from the server. When the client receives $\frac{\partial E}{\partial a^{(l)}}$, the client can generate $\frac{\partial E}{\partial z^{(l)}}$ by multiplying $\frac{\partial a^{(l)}}{\partial z^{(l)}}$, which is $g^{(l)}{'}{(z^{(l)})}$. Then the client is able to do backpropagation to the first hidden layer with \eqref{eq:1dconvbackward}. In backpropagation, there is no need for calculating gradients of input because it is not trainable. For example, the client does not have to calculate $\frac{\partial E}{\partial a^{(0)}}$, because the input is literally a untrainable parameter which is directly given by the user. However, from the server perspective, the server has to calculate the gradient of its input, $\frac{\partial E}{\partial a^{(l)}}$, because the client requires it to continue the backpropagation. Therefore, the model on the server part should be forced to calculate the gradient to the input level. To update weights, computing the gradient of weights in layer $i$ can be given as follows~\cite{kiranyaz2015real}:
\begin{equation}\label{eq:conv1dweight}
\frac{\partial E}{\partial w^{(i)}_{kj}} = \textup{Conv1D}(a_{k}^{(i-1)},\frac{\partial E}{\partial z^{(i)}_j})
\end{equation}


\end{document}